\pdfoutput=1

\documentclass[11pt, reqno]{amsart}
\usepackage[utf8]{inputenc}
\usepackage{amsmath}
\usepackage{comment}
\usepackage[dvipdfmx]{graphicx}
\usepackage{bbm}
\usepackage{hyperref}
\usepackage{caption}
\usepackage{tipa}
\usepackage{xspace}
\usepackage{bmpsize}

\usepackage[frozencache, cachedir=minted-cache]{minted}
\usepackage[margin=1.2in]{geometry}
\usepackage[htt]{hyphenat}
\AtBeginDocument{%
  }

\setminted[python]{bgcolor=bg,fontsize=\footnotesize,autogobble=True,python3=true}

\theoremstyle{definition}

\usepackage{subcaption}
\usepackage[utf8]{inputenc}
\usepackage{amsmath}
\usepackage{graphicx}
\usepackage{bbm}
\usepackage{hyperref}
\usepackage{caption}
\usepackage{tipa}
\usepackage{xspace}
\usepackage{minted}
\usepackage[numbers]{natbib}
\usepackage{graphics}
\usepackage{graphicx}
\usepackage{amsmath,amssymb,color}
\usepackage{colortbl}
\usepackage{multirow}
\usepackage{wrapfig}
\usepackage{xcolor}
\usepackage{wrapfig,lipsum,booktabs}
\usepackage{tabularx}

\newcommand*{\codeobj}[1]{\texttt{#1}\xspace}
\newcommand{\R}{\mathbb R}

\DeclareMathOperator*{\argmin}{argmin}

\AtBeginDocument{%
  }

\setminted[python]{bgcolor=bg,fontsize=\footnotesize,autogobble=True,python3=true}

\usepackage{todonotes}

\begin{document}
\definecolor{bg}{rgb}{0.95,0.95,0.95}

\title{Learning from landmarks, curves, surfaces, and shapes in Geomstats}
\author[Lu\'is F. Pereira, et al.]{Lu\'is F. Pereira, Alice Le Brigant, Adele Myers, Emmanuel Hartman, Amil Khan, Malik Tuerkoen, Trey Dold, Mengyang Gu, Pablo Su\'arez-Serrato, and Nina Miolane}
\date{}
\begin{abstract}
    We introduce the \texttt{shape} module of the Python package Geomstats to analyze shapes of objects represented as landmarks, curves and surfaces across fields of natural sciences and engineering. The \texttt{shape} module first implements widely used shape spaces, such as the Kendall shape space, as well as elastic spaces of discrete curves and surfaces. The \texttt{shape} module further implements the abstract mathematical structures of group actions, fiber bundles, quotient spaces and associated Riemannian metrics which allow users to build their own shape spaces. The Riemannian geometry tools enable users to compare, average, interpolate between shapes inside a given shape space. These essential operations can then be leveraged to perform statistics and machine learning on shape data. We present the object-oriented implementation of the \texttt{shape} module along with illustrative examples and show how it can be used to perform statistics and machine learning on shape spaces.
\end{abstract}

\maketitle

\section{Introduction}

Geomstats \cite{miolane2020geomstats} is an open-source Python package for statistics and learning from data that belong to \textit{manifolds}, \textit{i.e.}, to nonlinear generalizations of vector spaces.

\subsubsection*{Analyzing data on manifolds.} 
Manifolds arise in many applications. Hyperspheres
model directional data in molecular and protein biology \citep{Kent2005UsingStructure,Hamelryck2006,Villegas-Morcillo2021}. 
Hyperbolic spaces have recently gained interest to represent graph and hierarchical data across applications of computer vision~\citep{fang2023hyperbolic,mettes2023hyperbolic}. The manifold of symmetric positive-definite (SPD) matrices characterizes data from
diffusion tensor imaging (DTI) \citep{Pennec2006b,Yuan2012,PENNEC202075} and functional magnetic resonance imaging (fMRI)
\citep{Sporns2005TheBrain, FMRI2022}. The manifolds implemented in Geomstats come equipped with mathematical structures, such as Riemannian metrics, that allow users to process data belonging to them. For example, users can compute a distance between two data points on a manifold.
Geomstats also provides statistical learning algorithms that are compatible with the manifold structures, \textit{i.e.}, that can be applied to data belonging to any of the implemented manifolds. These algorithms are typically generalizations of traditional estimation, clustering, dimension reduction, classification and regression methods to nonlinear manifolds. For example, one algorithm called geodesic regression is the generalization of the traditional linear regression, but for data belonging to manifolds.

\subsubsection*{Analyzing shape data on shape spaces.} Shapes are a type of complex data that belong to nonlinear generalizations of vector spaces, called \textit{shape spaces}, including manifolds. Shape data are ubiquitous across fields of science and engineering. In molecular biology, the relationship between the shapes and functions of proteins is an active area of research. The statistical analysis of protein shapes, \textit{e.g.}, protein misfolding, helps understand illnesses such as Parkinson's disease \cite{JYLi2008}. 
Integrating advanced imaging techniques, such as cryogenic electron microscopy (cryo-EM), with image reconstruction tools, made it possible to determine  versatile 3D structures of  protein  at
near-atomic resolution \cite{li2013electron}. In evolutionary biology, the shape of monkey skulls, in combination with ecological and biogeographic data, is analyzed by paleontologists, gaining insights into evolutionary changes covering multiple geographies \cite{Elew2010}. 
In the medical realm, before an operation, orthopaedic surgeons analyze the shape of bones and then plan the surgery accordingly \cite{Darmante2014}. 
In computer vision, shape analysis of biological structures, for instance, seen as elements within shape manifolds, has  gained traction for a long time  \cite{Dryden1998,Younes2012Spaces}. 
In recent years, machine learning tools, such as U-Net \cite{ronneberger2015u}, have demonstrated their competitive performance in image segmentation, which can provide massive high-quality  shapes of different objects, such as cells \cite{stringer2021cellpose}, tissues and  organism \cite{siddique2021u}. 
The enhanced image segmentation methods hold great promise for probing disease progression such as fibrosis \cite{long2022mechanical}, by integrating cell shape, orientation and dynamics into the analysis  \cite{tisler2020analysis,luo2023molecular}. Similarly, in neuroimaging, researchers leverage the shape analysis of brain structure morphology, registered in MRI scans, advancing our comprehension of pathologies such as Alzheimer's disease \cite{Lorenzi2011}. Thus,  tools for processing and analyzing shape data can advance a broad range of scientific and engineering fields.

\subsubsection*{Implementing shape spaces in Geomstats.} 
To process and analyze  shape data, we introduce the \texttt{shape} module of the Python package Geomstats. 
Shape spaces in the \texttt{shape} module are implemented with the same design that drives the implementation of existing manifolds in Geomstats. Consequently, users can process and analyze shape data in the same way that they process and analyze manifold data. Specifically, the proposed \texttt{shape} module focuses on mathematical models of shapes where these are defined as the features of an \textit{object} that are invariant under certain \textit{transformations} \cite{Dryden2016,Small1996}. In other words, the shape of a set of landmarks, the shape of a curve, or the shape of a surface are defined as the remainder after we have filtered out the
position and the orientation of the object\textemdash or more generally after we have filtered out any transformations that do not change the shape of the object. Mathematically, this framework represents objects as elements of a fiber bundle equipped with the action of a group of transformations and shapes as elements of a quotient space that removes the group's action.

Examples of shapes and shape spaces that fall into this mathematical framework include the Kendall shape spaces \cite{Kendall1984}, quotient spaces from Procrustes analysis~\cite{Dryden1998}, and the elastic geometry of discrete curves~\citep{needham2020simplifying,mio2007shape,srivastava2007} and surfaces~\citep{hartman2023elastic,bauer2021numerical}. 

\subsubsection*{Related Works.} There are two main approaches to shape analysis: the extrinsic approach, where a shape is mapped to another by deforming the ambient space \cite{beg2005computing}, and the intrinsic approach, where the deformations are defined on the shapes themselves. Computational tools for the first approach have been proposed, see e.g.~\cite{gris2018sub} and the Python package Scikit-Shapes\footnote{\url{https://github.com/scikit-shapes/scikit-shapes}}. Here, we focus on the second, intrinsic approach. While implementations of some tools are available in C, Python and Matlab~\cite{Dryden2016,fdasrsf,LibSRVF2018,bauer2021numerical,hartman2023elastic},
to the best of our knowledge, there exists no wide-ranging open source Python implementation of intrinsic shape analysis, \textit{i.e.}, no implementation that tackles intrinsic metrics on shape spaces of landmarks sets, curves, and surfaces in a consistent mathematical framework. 



\subsubsection*{Contributions.}
This paper presents the \texttt{shape} module of Geomstats. Leveraging the fact that shape spaces bear crucial similarities thanks to their common quotient space structure, the module implements the differential geometry of object shapes and shape spaces, in particular: their group actions, fiber bundles, Riemannian and quotient structures. The implementation of object spaces and shape spaces is compatible with the main statistical learning algorithms of Geomstats' existing \texttt{learning} module. Thus, the proposed \texttt{shape} module unlocks the capacity to run learning algorithms on object data and shape data. As in the rest of Geomstats, the implementation of the \texttt{shape} modules is object-oriented and extensively unit-tested. All operations are vectorized for batch computation and support is provided for different execution backends — namely NumPy~\citep{numpy2020}, Autograd~\citep{maclaurin2015autograd}, and PyTorch~\citep{paszke2019pytorch}.

\subsubsection*{Outline.}
The rest of the paper is organized as follows.  Section~\ref{sec:background} reviews the mathematics of and implementation of the main package Geomstats. Then, Section~\ref{sec:module} introduces the structure of \texttt{shape} module, \textit{i.e.}, the abstract Python classes used to define objects and their shapes. Section~\ref{sec:catalogue} details the geometries of the concrete object and shape spaces implemented in the module, specifically objects and shapes of landmarks, curves and surfaces. This section includes code illustrations and examples of real-world use cases in the literature. Altogether, the proposed \texttt{shape} module represents the first comprehensive implementation of mathematical models of objects and their shapes in Python.

\section{Background: Differential Geometry in Geomstats}\label{sec:background}

In this section, we review the mathematics and design of the main package Geomstats required to understand the mathematics and design of the \texttt{shape} module. The package Geomstats implements concepts of differential geometry and Riemannian geometry following an object-oriented structure.
Abstract Python classes represent high-level mathematical concepts, such as \texttt{Manifold} (Subsection~\ref{subsec:manifold}), \texttt{Connection} (Subsection~\ref{subsec:connection}) and \texttt{RiemannianMetric} (Subsection~\ref{subsec:metric}). Their child Python classes then implement tangible manifolds, such as \texttt{Hypersphere}.

Here we give an informal introduction to Riemannian geometry. For more details, we refer the interested reader to standard textbooks such as \cite{do1992riemannian,lee2018introduction} and to \cite{guigui2023} for its original implementation in Geomstats.

\subsection{Manifold}\label{subsec:manifold}


A manifold is a space that locally resembles a vector space, without necessarily having its global flat structure. 
The simplest examples of manifolds are Euclidean spaces $\R^d$, or more generally vector spaces in finite dimensions, open sets of vector spaces 
and level sets of functions. 
A $d$-dimensional manifold $M$ admits a \textit{tangent space} $T_pM$ at each point $p\in M$ that is a $d$-dimensional vector space. 
The set of all tangent spaces to the manifold is called the tangent bundle and denotes $TM$. A manifold with a smooth differential structure is called a smooth manifold.



\subsubsection{Manifold in Geomstats}

The \codeobj{Manifold} abstract Python class implements the structure of a \textit{manifold}. Inheritance of Python classes allows us to specialize implementations. For example, \codeobj{VectorSpace}, \codeobj{OpenSet} and \codeobj{LevelSet} are abstract classes inheriting from the parent class \codeobj{Manifold}. Additionally, composition of Python classes allows us to combine already-existing structures. For example, \codeobj{ProductManifold} is a Python class that represents a manifold created as the product of two or more existing manifolds. Similarly, \codeobj{NFoldManifold} is a class that represents a manifold created as the product of a base manifold repeated $n$ times.

The parent class \codeobj{Manifold} and its child classes contain methods that allow users to process data points on manifolds. For example, a user can verify that a given data point indeed belongs to the manifold via the \codeobj{belongs()} method and that a given input is a tangent vector to the manifold at a given base point via \codeobj{is\_tangent()}. Additionally, a user can generate random data points and tangent vectors to the manifold with the methods \codeobj{random\_point()} and  \codeobj{random\_tangent\_vec()} respectively. The latter two methods will be particularly relevant for the unit-testing framework.

\subsection{Connection}\label{subsec:connection}

An \textit{affine connection} is a mathematical structure that allows us to define the generalization of straight lines, addition and subtraction to nonlinear manifolds, respectively called geodesics, exponential map and logarithm map. To this end, a connection allows us to take derivatives of vector fields, \textit{i.e.}, mappings $V:M\rightarrow TM$ that associate to each point $p$ a tangent vector $V(p)\in T_pM$. 
The derivative induced by the connection $\nabla$ is referred to as \textit{covariant derivative}. In a local coordinate system along the manifold, the coefficients of the connection are called the Christoffel symbols. Mathematical details on connections, covariant derivatives, and Christoffel symbols can be found in \cite{do1992riemannian}.

\subsubsection{Geodesics}

Consider $t \mapsto \gamma(t)$ a curve on $M$, parameterized by time $t$. Its velocity $t \mapsto \dot\gamma(t)$ is a vector field along $\gamma$, \textit{i.e.}, $\dot\gamma(t)\in T_{\gamma(t)}M$ for all $t$. The acceleration of a curve is, by definition, the covariant derivative of this velocity field with respect to the affine connection $\nabla$. A curve $\gamma$ of zero acceleration 
\begin{equation}
    \label{general_geod_eq}
    \nabla_{\dot\gamma}\dot\gamma=0,
\end{equation}
is called a \textit{$\nabla$-geodesic}. Geodesics are the manifolds counterparts of vector spaces' straight lines, \textit{i.e.}, their second derivative vanishes (in the sense of connections). 
 Equation~\eqref{general_geod_eq} becomes a system of ordinary differential equations (ODEs) that can be solved to find geodesics. 

\subsubsection{Exponential and logarithm maps}
Existence results for solutions of ODEs allow us to define geodesics starting at a point $p$ with velocity $v\in T_pM$ for times $t$ in a neighborhood of zero, or equivalently for all time $t\in[0,1]$ but for tangent vectors $v$ of small norm. The \textit{exponential map} at $p\in M$ associates to any $v\in T_pM$ of sufficiently small norm the end point $\gamma(1)$ of a geodesic $\gamma$ starting from $p$ with velocity $v$:
$$\exp_p(v)=\gamma(1), \quad \text{where} \begin{cases} 
\gamma \text{ is a geodesic}, \\ 
\gamma(0) = p, \, \dot\gamma(0) = v.
\end{cases}$$
The map $\exp_p$ is a diffeomorphism in a neighborhood of $0$ in $T_pM$, and its inverse $\log_p \equiv \exp_{p}^{-1}$ defines the \textit{logarithm map}. The logarithm map then associates to any point $q$ the velocity $v\in T_pM$ necessary to get to $q$ when departing from $p$:
$$\log_{p}(q)=v \quad\text{where}\quad  \exp_p(v)=q.$$
The exponential and logarithm maps can be seen as generalizations of the Euclidean addition and subtraction to nonlinear manifolds. Indeed, the exponential map \textit{adds} a tangent vector to a point which outputs a point. Similarly, the logarithm map \textit{subtracts} two points and outputs a tangent vector. Geodesics can in turn be expressed in terms of the exponential map: if a geodesic $\gamma$ verifies $\gamma(0)=p$ and $\dot\gamma(0)=v$, then $\gamma(t) = \exp_{p}(tv)$.

\subsubsection{Connection in Geomstats}

The \codeobj{Connection} class implements the structure of an \textit{affine connection}. This class first contains the \codeobj{christoffels()} method that implements the Christoffel symbols defining the connection. Just as a connection allows us to define geodesic equation and geodesics, the \texttt{Connection} class contains the \codeobj{geodesic\_equation()} method implementing Equation~\eqref{general_geod_eq}, as well as the \codeobj{geodesic()} method computing geodesics either from initial conditions $\gamma(0)$, $\dot\gamma(0)$ or from boundary conditions $\gamma(0)$, $\gamma(1)$. Similarly, we find the
\codeobj{exp()} and \codeobj{log()} methods for exponential and logarithm maps, respectively. We refer to \cite{guigui2023} for additional details on the \codeobj{Connection} class.

\subsection{Riemannian metric}\label{subsec:metric}

A \textit{Riemannian metric} is a collection of inner products $(\langle\cdot,\cdot\rangle_p)_{p\in M}$ defined on the tangent spaces of a manifold $M$, that depend on the base point $p\in M$ and vary smoothly with respect to it. 

\subsubsection{Levi-Civita Connection} Given a Riemannian metric there exists a unique affine connection, called the \textit{Levi-Civita connection}, which is the only affine connection that is symmetric and compatible with the metric, \textit{i.e.}, that verifies
\begin{align*}
    UV-VU = \nabla_UV - \nabla_VU\\
    U\langle V,W\rangle = \langle \nabla_UV, W\rangle + \langle V,\nabla_UW\rangle
\end{align*}
for all vector fields $U,V,W$. The geodesics of a Riemannian manifold are those of its Levi-Civita connection.

\subsubsection{Geodesic Distance}
The \textit{geodesic distance} induced by the Riemannian metric is defined as the length of the shortest curve joining two points $p, q\in M$. Here, the length of a (piecewise) smooth curve $\gamma:[0,1]\rightarrow M$ is computed by integrating the norm of its velocity using the norm induced by the Riemannian metric:
$$d(p, q)=\inf_{\gamma;\gamma(0)=p, \gamma(1)=q}L(\gamma),\quad \text{where}\quad L(\gamma)=\int_0^1 || \dot\gamma(t) ||_{\gamma(t)} dt.$$

In a Riemannian manifold, geodesics extend another property of straight lines: they are locally length-minimizing. 
In a geodesically complete manifold, any pair of points can be linked by a minimizing geodesic, not necessarily unique, and the distance between them can be computed using the logarithm map, written for all $p, q$ in $M$ as follows: 
\begin{align*}
    \quad d(p, q) = ||\log_p(q)||_p.
\end{align*}



\subsubsection{Riemannian metric in Geomstats}

The abstract class \codeobj{RiemannianMetric} implements the structure of a Riemannian metric. It is a child class of \codeobj{Connection} and inherits all its methods, including \codeobj{geodesic()}, \codeobj{exp()} and \codeobj{log()}. The class \codeobj{RiemannianMetric} overwrites the \codeobj{Connection} class' method \codeobj{christoffels()} and computes the Christoffel symbols using derivatives of the metric by using automatic differentiation. The geodesics, by the compatibility property, have velocity of constant norm, \textit{i.e.}, are parametrized by arc length. The \codeobj{dist()} method implements the geodesic distance induced by the Riemannian metric.

Any \codeobj{Manifold} can be equipped with a Riemannian metric through a \codeobj{equip\_with\_metric()} method, \textit{i.e.}, it sets \codeobj{metric} as an attribute of \codeobj{Manifold}. In other words, \codeobj{equip\_with\_metric()} transforms a smooth manifold $M$ into a Riemannian manifold $(M, g)$. This method is particularly important since a manifold may be equipped with different metrics.

\section{The Shape Module of Geomstats}\label{sec:module}

\begin{figure}[h]
\centering
\includegraphics[width=0.9\linewidth]{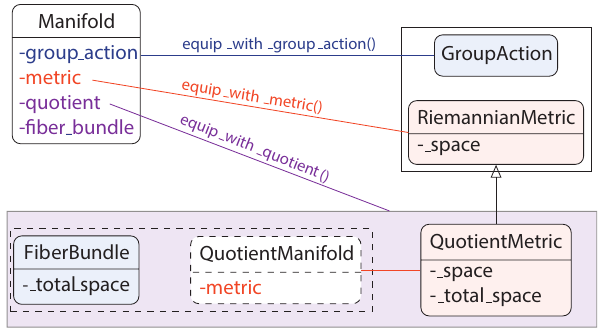}
\caption{\textbf{Abstract Python classes of the \texttt{shape} module in Geomstats.} A manifold can be equipped with a group action and a Riemannian metric through the methods \texttt{equip\_with\_group\_action()} and \texttt{equip\_with\_metric()}. When those are compatible, a quotient structure can be put on the manifold through the method \texttt{equip\_with\_quotient()}. The quotient structure consists of a fiber bundle, and a quotient manifold equipped with a quotient metric, which inherits from the class \texttt{RiemannianMetric} (arrow represents inheritance). \codeobj{QuotientManifold} is dashed because it may be the same class (but different instance) as \codeobj{Manifold} when points are represented via their representatives. The variables listed under each Python class represent the main attributes of an object instantiated from this class.}
\label{fig:quotient-structure}
\end{figure}

We can now introduce the \texttt{shape} module of Geomstats. The architecture of the module follows an object-oriented design, consistent with the design of the main package Geomstats. Abstract Python classes represent high-level mathematical concepts, such as as \texttt{FiberBundle} (Subsection~\ref{subsec:bundles}), \texttt{GroupAction} (Subsection~\ref{subsec:actions}), and \texttt{QuotientMetric} (Subsection~\ref{subsec:quotient}) ~| summarized in Figure~\ref{fig:quotient-structure}. In this section, we review the  differential geometry of shape spaces which motivates the architecture of the \texttt{shape} module. The module is available in the three backends of Geomstats: NumPy, Autograd and PyTorch. Additionally, the code is extensively unit-tested, documented and included in the continuous integration pipeline and documentation website of the Geomstats library.

\subsection{Fiber Bundles}\label{subsec:bundles}

In shape analysis, a space of objects can be represented through the mathematical structure of a fiber bundle, which we introduce here.

Let $M,B$ and $F$ be smooth manifolds and suppose that $\pi: M \rightarrow B$ is smooth. The triple $(\pi, M, B)$ is called \textit{fiber bundle with total space $M$, base space $B$ and fiber $F$} if: 

\begin{itemize}
    \item[$i)$] $\pi$ is surjective;
    \item[$ii)$] there exists an open cover $(U_i)_{i \in I}$ of $B$ and diffeomorphisms
    \begin{align*}
        h_i: \pi^{-1}(U_i) \rightarrow U_i \times F
    \end{align*}
    such that $h_i(\pi^{-1}(p)) = \{p\} \times F.$
\end{itemize}

Informally, the total space $M$ of a fiber bundle \textit{locally} looks like a product of the base $B$ with the fiber $F$. In the context of shape analysis, the total space $M$ will typically represent a space of objects, whereas the space $B$ will represent a space of these objects' shapes. The map $\pi$ represents the extraction of a shape from an object, and $\pi^{-1}(p)$ represents all objects that have shape $p$. 

\subsubsection{Fiber bundles in the \texttt{shape} module} The Python class \codeobj{FiberBundle} implements the differential geometry of fiber bundles in the \texttt{shape} module~| see Figure~\ref{fig:quotient-structure}. In particular, it implements methods that transform points from the base space $B$ to the total space $M$, and vice-versa. These methods are summarized in Table~\ref{tab:fiber-bundle}, and are discussed in the next subsections in more details. 

\begin{table}[]
    \centering
    \begin{tabular}{c|l} 
        Method & Description\\\hline
        \codeobj{riemannian\_submersion()}& $M \to B$ \\
        \codeobj{lift()}& $B \to M$\\
        \codeobj{tangent\_riemannian\_submersion()}& $T_p M \to T_{\pi(p)} B$\\
        \codeobj{horizontal\_lift()}& $T_{\pi(p)} B \to T_p M$ \\
        \codeobj{horizontal\_projection()}& $T_p M \to H_p $\\
        \codeobj{vertical\_projection()} & $T_pM \to V_p$\\
        \codeobj{align()} & $ G\cdot p \to G\cdot p$
    \end{tabular}
    \caption{\textbf{Main methods for the Python class \codeobj{FiberBundle}}. Abbreviations: $M$: total space, $B$: base space, $\pi$: Riemannian submersion, $T_p M$: tangent space at $p \in M$, $H_p$: horizontal subspace at $p \in M$, $V_p$: vertical tangent space at $p \in M$, $G$: group acting on $M$.}
    \label{tab:fiber-bundle}
\end{table}

\subsection{Group Actions}\label{subsec:actions}

In shape analysis, objects can be transformed by translations, rotations, and other geometric transformations. These ideas are mathematically formulated through the concepts of groups and group actions ~| which we introduce now. A group is a pair $(G,\ \circ)$ where $G$ is a set and $\circ:G\times G\mapsto G$ is an associative multiplication with an identity element $\text{id}\in G$ and such that any $\phi\in G$ has an inverse $\phi^{-1}\in G$.
A Lie group $(G, \circ)$ is a group that is also equipped with a smooth manifold structure, such that both the group action $\circ : G \times G \rightarrow G$ and map $G\rightarrow G,$ $\phi \mapsto \phi^{-1}$ are smooth. 

A Lie group $(G,\circ)$ \textit{acts} on a smooth manifold $M$, if there exists a smooth map $\cdot : G \times M \rightarrow M$ such that
\begin{align*}
   \phi_1 \cdot (\phi_2\cdot p) = (\phi_1 \circ \phi_2) \cdot  p \quad \textup{for all }\phi_1,\phi_2 \in G, \, \textup{and }p \in M.
\end{align*}

The Lie group $G$ is said to act \textit{freely} on $M$ if $ \phi \cdot p \neq p $ for all $\phi \neq \textup{id}\in G$ and $p\in M$ and is said to act \textit{properly}
 on $M$ if for any compact $K \subset M$ the set 
 \begin{align*}
     G_K = \{ \phi \in G \, |\, \phi K \cap K \neq \emptyset\}
 \end{align*}
 is relatively compact in $G$ (i.e. $\overline G_K \subset G$ is compact), where $\phi K$ is the image of the map $\phi:K \rightarrow G$, $p \mapsto \phi \cdot p.$

\subsubsection{Group actions in the \texttt{shape} module} 
The Python class \codeobj{LieGroup} of Geomstats implements the mathematical structure of Lie groups. This class contains for example the method \texttt{compose()} that represents the composition operation $\circ$ of the group. In the \texttt{shape} module specifically, the Python class \codeobj{GroupAction} implements the action of a group element $\phi$ on a point $p$ of the manifold~| see Figure~\ref{fig:quotient-structure}. 
Analogously to equipping a manifold with a metric, \codeobj{equip\_with\_group\_action()} method allows to endow a \codeobj{Manifold} with a group action.

\begin{figure}[h]    \includegraphics[width=0.7\linewidth]{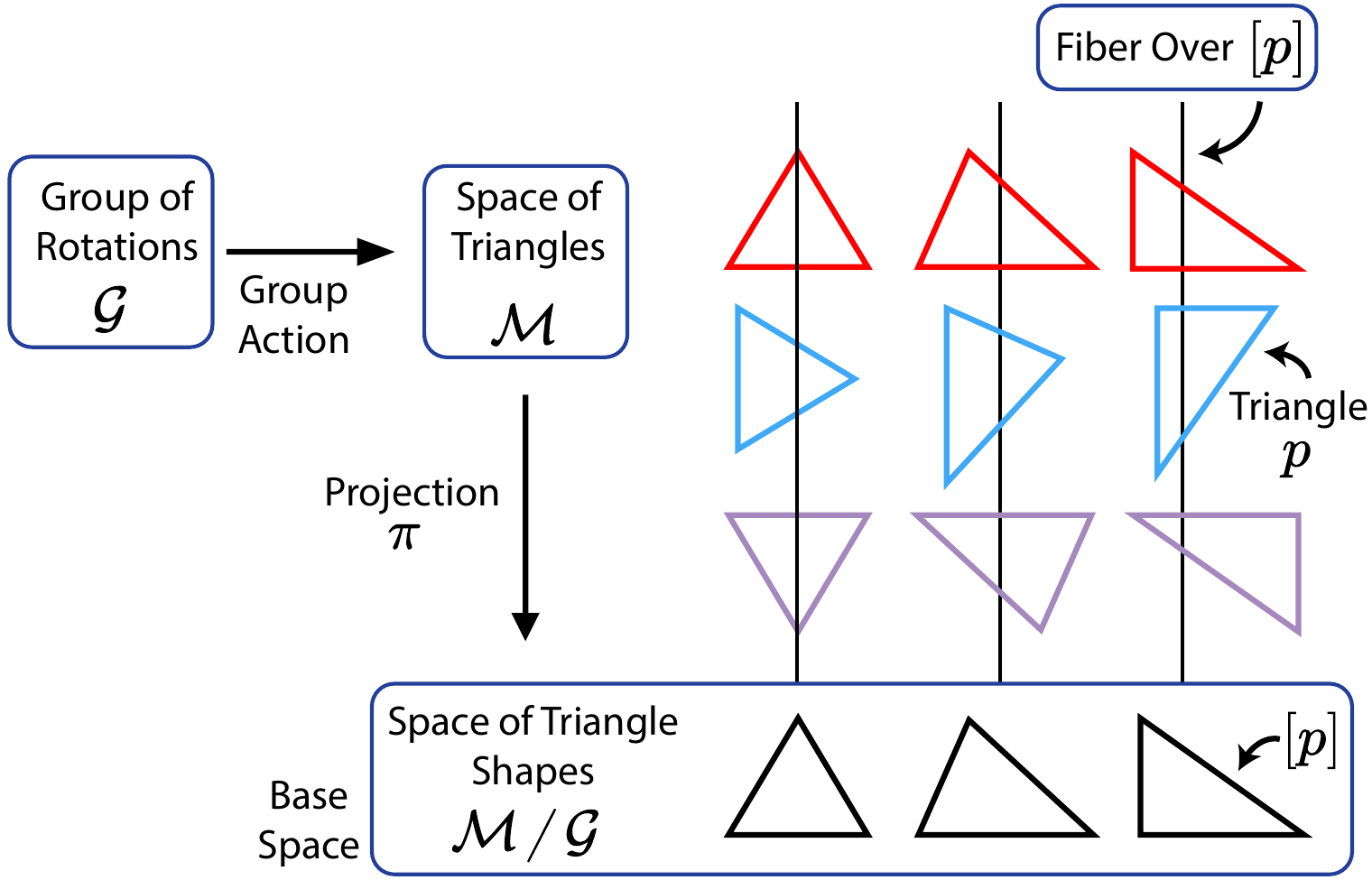}
    \caption{\textbf{Example of fiber bundle induced by a group action.} The group of rotations $G$ acting on the space of triangles $M$ induces a fiber bundle with total space $M$, base space $M/G$ and fiber $G$. The fiber over $[p]$ is the orbit $p\cdot G$ composed of all triangles that can be obtained by rotating $p$. They all project on the same shape $[p]\in M/G$.}
    \label{fig:fiber_bundle}
\end{figure}

 \subsection{Quotient Space}\label{subsec:quotient}

In shape analysis, the shapes of objects can be represented by points in a quotient space, defined next. Consider a group $G$ that acts on a point $p$ of a manifold $M$: this will reach a set of points $G \cdot p$, called the \textit{orbit} of $p \in M$ through the action of $G$. This orbit defines an equivalence class, that we denote $[p]$. By definition, the set of equivalence classes is called a \textit{quotient space} and denoted $M / G$. A group $G$ acting on $M$ defines a fiber bundle with total space $M$, base space $M/G$, and fiber $G$. Each orbit $G\cdot p$ is homeomorphic to $G$, and is therefore called the fiber over $[p]$.

For example, as shown in Figure~\ref{fig:fiber_bundle},
$M$ can be a set of objects such as a set of triangles and $G$ can be the group of rotations and translations. In this scenario, $p$ is a triangle, and $[p]$ represents all triangles that are obtained from $p$ by rotations and translations. Intuitively, $[p]$ defines a notion of shape: the shape of a triangle $p$ is defined as all the triangles that have the same shape as $p$.


 

\subsubsection{Quotient Space in the \texttt{shape} module}

In order to compute with shape data, we need to compute with points in a quotient space. Points in a quotient space $M / G$ are equivalence classes. In general, there is no obvious way to explicitly implement an equivalence class in the computer. Consequently, we choose to (implicitly) represent an equivalence class $[p]$ of $M / G$ as an element $p \in [p]$ of the total space $M$. In doing so, $p$ is called the \textit{representative} of the class. All quotient spaces presented in the next subsection represent points via representatives in the total space $M$.

When points on the quotient space are represented via representatives, we can use the fundamental operations of Riemannian geometry of the total space to compute in the quotient space.

\subsection{Quotient Metric} {\label{sec:quotient_dist}}

We need a notion of compatibility between the group action and the metric structure in order to define a distance on the quotient space. We say that $G$ \textit{acts 
isometrically 
on a Riemannian manifold} $(M, g)$ if for the action $\cdot : G\times M \rightarrow M,$ one has that  
\begin{align*}
    d(p,q) = d(\phi \cdot p, \phi\cdot q) \quad \textup{for } p,q\in M \quad \textup{and }\phi \in G.
\end{align*}
Equivalently, we say that the metric is $G$-invariant, or that the metric and the group action are compatible. When the Riemannian metric and the group action by $G$ are compatible, we obtain a definition of distance on the quotient space $M /G: $
\begin{equation}\label{eq:quotient-dist}
    d_{M/G}([p], [q]) = \inf_{\phi \in G} d_M( p, \phi \cdot q).
\end{equation}
In our example with triangles, the distance between the shape $[p]$ of the triangle $p$ and the shape $[q]$ of the triangle $q$ is the minimum of the distances in object space between the triangle $p$ and all possible rotations and translations of the triangle $q$. 

If $G$ acts freely and properly on $M$, the distance \eqref{eq:quotient-dist} is induced by a Riemannian metric on the quotient space, such that the projection $\pi$ is a Riemannian submersion, \textit{i.e.}, an orthogonal projection along the fibers. The space $V_p:=\ker d\pi_p$ is called the \textit{vertical subspace} of $T_pM$ and is composed of vectors of $T_pM$ that are tangent to the fibers; following such vectors amounts to staying in the same equivalence class. In the context of shape analysis, vertical vectors correspond to infinitesimal perturbations of landmarks, curves or surfaces that do not change the shape. The orthogonal complement of the vertical subspace in $T_pM$ is called the \textit{horizontal subspace} and denoted $H_p$. 

\subsubsection{Quotient Metric in the \texttt{shape} module} 

The Python class \codeobj{QuotientMetric} inherits from \codeobj{RiemannianMetric}, overriding most of the methods with general quotient space-specific implementations that take advantage of the methods available in \codeobj{FiberBundle} and the total space metric.

When a total space $M$ is equipped with a metric and a group action, we can use the method \codeobj{equip\_with\_quotient()} to create a quotient space (see Figure~\ref{fig:quotient-structure}). This method instantiates a base space equipped with a quotient metric and equips the total space with a fiber bundle. A quotient registry maps total spaces, metrics and group actions to base spaces, fiber bundles and quotient metrics. The function \codeobj{register\_quotient()} allows users to register their own quotient structures.

\subsection{Alignment}{\label{sec:alignment}}

The definition of distance in the quotient space $M/G$ in Eq.~\eqref{eq:quotient-dist} relies on a notion of alignment. Indeed, in this equation, we pick the best representative of an orbit $G\cdot q = [q]$ with respect to the point $p$, \textit{i.e.}, we pick the point on the orbit $[q]$ for which the (total space) distance to $p$ is minimized. Computing a distance between two points in $M/G$ relies on our ability to run an alignment procedure.

On the one hand, the alignment procedure is straightforward when the group $G$ acting on the manifold is finite-dimensional. In this case, an explicit representation of the group elements exists. Consequently, the alignment can be solved by a gradient-based minimization of the (squared) distance to $p$ over the group. To this aim, we can leverage the automatic differentiation capability of Geomstats; though closed-form solutions are implemented for particular cases (\textit{e.g.}, rotations acting on matrices).

On the other hand, the alignment procedure is very challenging when the group is infinite-dimensional. Such an infinite-dimensional group arises in the context of the shape spaces of curves and surfaces in the next section: the group of reparametrizations.
A first solution consists in discretizing the group \citep{laga2017, su2020surfaces} and/or the action of the group \citep{mio2007shape}. However, this approach is limited in scope (\textit{e.g.}, it requires surfaces to be of genus-zero in the case of shapes of surfaces) or leads to computationally expensive algorithms (\textit{e.g.}, dynamic programming-based algorithm to align discrete curves). 
There exists an alternative solution that avoids having an explicit parameterization of the group. In this approach, we can formulate the geodesic boundary value problem on the quotient space as a relaxed numerical optimization problem involving an orbit membership enforcing term $\Gamma\left(\gamma(1), q\right)$ \citep{bauer2018}, \textit{i.e.}, the geodesic $\gamma$ between a point $p$ and the (approximate) orbit of a point $q$ is the path resulting from
\begin{equation}\label{eq:relaxed-bvp}
\argmin _{\gamma \in \mathcal{P}_{p}} E(\gamma) +\lambda \Gamma\left(\gamma(1), q\right)
\end{equation}
where $\mathcal{P}_{p}$ is the space of paths starting at $p$, and $E(\gamma)$ is the Riemannian energy of path $\gamma$. 
This approach is closely related to the inexact formulation of the LDDMM (Large Diffeomorphic Deformation Metric Mapping) framework, where the relaxation term is known as fidelity or data attachment term \citep{charon2020fidelity}. It has been introduced in the setting of discrete curves \citep{bauer2018}, and later extended to the setting of discrete surfaces \citep{bauer2021numerical}. This formulation admits a symmetrized version:
\begin{equation}\label{eq:sym-relaxed-bvp}
\argmin _{\gamma\in \mathcal{P}} E(\gamma) +\lambda_0 \Gamma\left(\gamma(0), p\right) + \lambda_1 \Gamma\left(\gamma(1), q\right)
\end{equation}

\subsubsection{Alignment in the \texttt{shape} module} 



The numerical procedure associated with the alignment approach is implemented in the \texttt{align()} method of the \texttt{FiberBundle} class, see Table~\ref{tab:fiber-bundle}. An abstract Python class \codeobj{AlignerAlgorithm} can be passed to \codeobj{FiberBundle} to implement \texttt{align()} in the general case, when no closed-form exists. If there exists a closed-form solution to the alignment problem, then it is implemented by overriding the \codeobj{align()} method of \codeobj{FiberBundle}.

Together with the the horizontal and vertical projection of a tangent vector, the alignment procedure is central to all computations in the quotient space. We already saw how the distance between two points amounts to aligning them and computing the distance using the total space metric. Additionally, the exponential map of a tangent vector at a base point amounts to horizontally projecting the tangent vector and shooting using the total space metric. Similarly, the logarithm map of a point at a base point amounts to aligning point to base point and taking the logarithm map of the aligned point at base point using the total space metric.

\subsection{Remark: Combining Group Actions}

Central to shape spaces is the ability to easily combine group actions, as the notion of ``shape'' is application-dependent. For example, we may want to combine the action of rotations and the action of reparametrizations when we consider shapes of curves and surfaces in the next section.

Assume we have $k$ groups acting on a manifold, and we know how to align points with respect to each individual group action. A general alignment procedure consists of a $k$-step iterative process, where at each step the alignment with respect to the action $0 \le i \le k-1$ is performed \citep{srivastava2016book}.

Combination of group actions is achieved through the use of a tuple of group actions. In this case, an alternating alignment algorithm is used by default (\codeobj{AlternatingAligner}).


 \section{Landmarks, Curves, Surfaces and Their Shapes in the \texttt{shape} Module}\label{sec:catalogue}

This section details the concrete object spaces and shape spaces that we implement in the \codeobj{shape} module, specifically: landmark sets, curves, surfaces spaces and their corresponding shape spaces. As such, this section also provides a comprehensive review of shape analysis. Each subsection further showcases code snippets using each shape space to demonstrate the diversity of use cases of the proposed \codeobj{shape} module.

\subsection{Landmark Sets}

\subsubsection{Pre-shape space} 
\begin{figure*}[h]
  \centering
  \includegraphics[width=\textwidth]{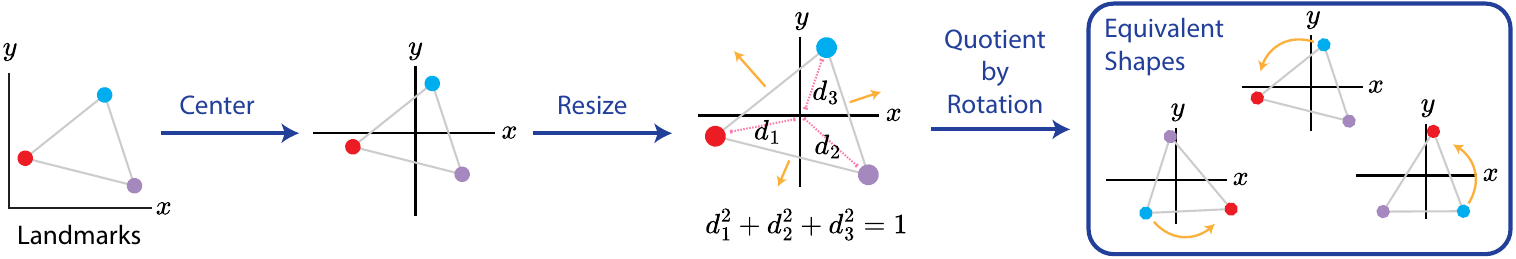}
    \caption{\textbf{Landmark sets.} Consider an object defined as a set of 3 landmarks in 2D (left). We transform this object into a shape, \textit{i.e.}, into a point in Kendall shape space via the following operations: center the landmark set by placing its barycenter at the origin, resize the landmark set so that its `size' equals 1, and then quotient by the group of rotations such that the difference in rotation will not contribute to distance between points in Kendall shape space. ``Equivalent shapes'' in the figure all correspond to the same point in Kendall shape space}.
    \label{fig:kendall-implementation}
\end{figure*}

Consider the space of $k$ landmarks of $\mathbb{R}^d$. A data point of this space is $p \in M(k, d)$  where $M(k, d)$ is the space of real $k \times d$ matrices. Changing the position and size of a landmark set $p$ do not change its shape. After removing translations ($p_i - \bar{p}$, where $\bar{p}$ is the barycenter of $p$) and scaling (by dividing by the Frobenius norm here denoted by $|| \cdot ||$), we obtain the \emph{pre-shape space} (see, \textit{e.g.}, \citep{Dryden2016, guigui2021}) defined by
$$
\mathcal{S}(k, d)=\left\{p \in M(k, d) \mid \sum_{i=1}^k p_i=0,\,  || p || = 1 \right\}.
$$
The Frobenius metric on $M(k, d)$ induces a metric on $\mathcal{S}(k, d)$ by the embedding $i :\mathcal{S}(k, d) \rightarrow M(k, d)$, known as the spherical Procrustes metric. This metric makes $\mathcal{S}(k, d)$ isometric to the round sphere $\mathbb S^{d (k-1) - 1}\subset \mathbb R^{d(k-1)}.$ The left side of Figure~\ref{fig:kendall-implementation} illustrates the operation of centering and resizing that converts an original landmark set of $k=3$ landmarks in $d=2$ dimensions into an element of the pre-shape space.

\subsubsection{Kendall shape space} Rotating a set of landmarks does not change its shape. Accordingly, the \emph{Kendall shape space}, denoted by $\Sigma(k, d),$ is the quotient space induced by the action of the group of rotations $\mathrm{SO}(d)$ on $\mathcal{S}(k, d)$:
\begin{equation}
    \Sigma(k, d) = \mathcal{S}(k, d) / \mathrm{SO}(d).
\end{equation}
The quotient metric on $\Sigma(k, d)$ induced from this group action and quotient structure is the so-called \emph{Kendall metric}. Note that this action is smooth, proper but not free when $d \ge 3$, which is why the quotient space has singularities (see, \textit{e.g.}, \cite{Dryden2016, guigui2021}). In other words, this space is a Riemannian manifold when $d=2$ (complex projective space) and a stratified space when $d \ge 3$ \citep{Dryden2016}. Away from singularities, the canonical projection $\pi: \mathcal S(k, d) \rightarrow \Sigma(k, d)$ is a Riemannian submersion. The full preprocessing evolution which takes unprocessed landmarks to a point in Kendall shape space is shown in Fig.~\ref{fig:kendall-implementation}, where the right side corresponds to the quotient by rotations.



The fiber bundle structure associated with Kendall shape spaces is well known. In particular, the vertical and horizontal subspaces can be computed explicitly.
The vertical subspace at $p\in\mathcal{S}(k, d)$ is given by 
$$V_p=\{p A^T\mid A \in \operatorname{Skew}(d)\},
$$
where $\operatorname{Skew}(d)$ denotes the space of skew-symmetric $d\times d$ matrices. The horizontal subspace is given by 
$$
\begin{aligned}
H_p & =\left\{w \in T_p \mathcal{S}(k, d)\mid \operatorname{Tr}\left(A p^T w\right)=0, \, \forall A \in \operatorname{Skew}(d)\right\} \\
& =\left\{w \in T_p \mathcal{S}(k, d) \mid p^T w \in \operatorname{Sym}(d)\right\},
\end{aligned}
$$
where $\operatorname{Sym}(d)$ is the space of symmetric $d \times d$ matrices.
The vertical component of a tangent vector can be computed as $\operatorname{Ver}_p(w)= p A^T$ \citep{yazdani2020}, where $A$ solves the Sylvester equation:
$$
A p^T p+p^T p A=w^T p-p^T w.
$$

Given a geodesic $\gamma$ in $\mathcal S(k, d)$ such that $\dot{\gamma}(0)$ is a horizontal vector, then   $\dot{\gamma}(t)$ is horizontal for all $t$, and in particular, $\pi \circ \gamma$ is a geodesic of 
$\Sigma(k, d)$. We leverage these facts in our implementation of the Kendall shape spaces in the \texttt{shape} module.

\subsubsection{Aligning landmark sets}

Algorithms to align landmark sets are central to perform computations on the Kendall shape space. Here, we align landmark sets using the finite-dimensional group of rotations $SO(d)$. This approach is related to Procrustes analysis literature, where landmark sets are aligned using different groups of transformations such as general invertible affine transformations. The alignment strategy implemented in the \texttt{shape} module to find optimal rotation aligning a given set of landmarks onto another one uses a traditional approach leveraging SVD.

\subsubsection{Landmark sets and their shapes in the \texttt{shape} module}

\begin{figure}[h]
\centering
\includegraphics[width=0.9\linewidth]{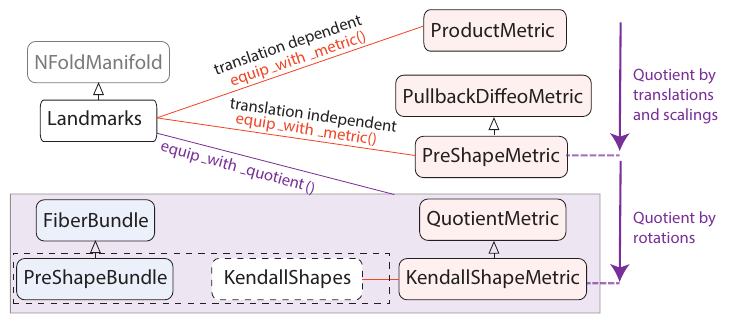}
\caption{\textbf{Landmark sets and their shapes in the \texttt{shape} module of Geomstats.} The Python class \texttt{Landmarks} (left) represents a space of objects, a pre-shape space, or a Kendall shape space depending on the Riemannian metric that equips it: with the product Euclidean metric, the pre-shape metric, or the Kendall shape metric, respectively (right). Each of these metrics quotients specific group actions: translations, scalings, and rotations, as shown by the purple arrows (right). We indicate the abstract Python classes from which each class inherits, with empty black arrows representing inheritance.}
\label{fig:kendall-structure}
\end{figure}

The \codeobj{Landmarks} class implements the space $M(k, d)$ of matrices representing $k$ landmarks in $d$ dimensions, using a product manifold $M = \mathbb{R}^d \times \cdots \times \mathbb{R}^d$ in the Python class \codeobj{Landmarks} which inherits from the \codeobj{NFoldManifold} class. To instantiate a space of triangles in 2D, users can employ the following code:

\begin{minted}{python}
from geomstats.geometry.euclidean import Euclidean
from geomstats.geometry.landmarks import Landmarks

euclidean = Euclidean(dim=2)
landmarks_space = Landmarks(k_landmarks=3, ambient_manifold=euclidean)
\end{minted}

The \codeobj{PreShapeSpace} class, which inherits from the abstract class \codeobj{LevelSet}, implements the pre-shape space $\mathcal{S}(k, d)$ (see Figure~\ref{fig:kendall-structure} for details). Points are represented by $k \times d$ matrices, \textit{i.e.}, we scale centered landmark coordinates $X_C$ by the Frobenius norm ($x_C / ||x_C ||$) to obtain valid pre-shape space points. Alternatively, \textit{e.g.}, Helmertized landmark coordinates, where points are represented by $(k-1) \times d$ matrices, could have been used \cite{Dryden2016}. The \codeobj{PreShapeMetric} class implements the spherical Procrustes metric. These implementations take advantage of the round sphere $S^{d k -1}$ implementation, as points only need to be flattened to directly apply it. 

The \codeobj{KendallShapeMetric} class inherits from the class \codeobj{QuotientMetric}. Similarly, the \codeobj{PreShapeBundle} class inherits from the \codeobj{FiberBundle} class. Both classes implement the quotient structures required for the Kendall shape space. We note that computations related to the affine connection, such as the so-called parallel transport, on this space have been implemented by the algorithm introduced in \citep{guigui2021}. 

In the \texttt{shape} module, a pre-shape space $\mathcal{S}(k, d)$ of triangles in 2D, \textit{i.e.}, with $k=3$ and $d=2$, is instantiated with the following code:

\begin{minted}{python}
from geomstats.geometry.pre_shape import PreShapeSpace

preshape = PreShapeSpace(k_landmarks=3, ambient_dim=2)
\end{minted}

Next, in order to create a Kendall shape space, we equip the pre-shape space with the action of the group of rotations $SO(d)$ which we quotient out. The resulting quotient space is the Kendall shape space.

\begin{minted}{python}
preshape.equip_with_group_action("rotations")
preshape.equip_with_quotient()

kendall_shape = preshape.quotient
kendall_metric = kendall_shape.metric
\end{minted}

We observe that the Kendall shape space comes equipped with its Kendall shape metric. The user can access essential computations of Riemannian geometry, such as \texttt{kendall\_metric.exp()}, \texttt{kendall\_metric.log()} and \texttt{kendall\_metric.geodesic()}.

\subsubsection{Use case: Regression on landmark set shapes} 
\begin{figure}[ht]
    \centering
    \includegraphics[width=\textwidth]{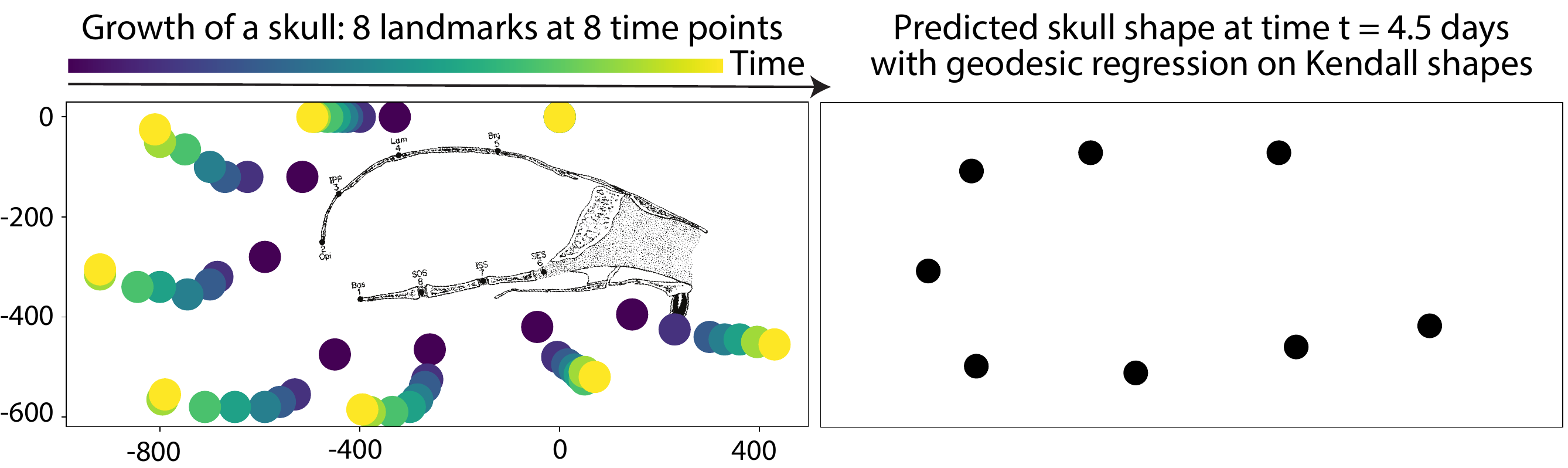}
    \caption{\textbf{Prediction of rat's skull shapes using the \texttt{shape} module of Geomstats}. Each rat skull is represented by a set of $k=8$ landmarks in $d=2$ dimensions. Left: Skull shape of rats at 8 time points, at ages of 7, 14, 21, 30, 40, 60, 90, and 150 days (different times shown by different color) \cite{Bookstein1991}. A anatomical drawing of a rat skull is shown under the scatter plot for reference. Right: Predicted shape of the rat skull after 4.5 days after geodesic regression on Kendall shape space.}
    \label{fig:rat-skulls}
\end{figure}

We show how to perform machine learning on landmark sets shapes implemented in the \texttt{shape} module using a regression method available in the main Geomstats package. Specifically, we demonstrate the use of geodesic regression \citep{Fletcher2013} to estimate the shape of a rat's skull as a function of time. In this example, we use Vilmann's rat calvaria (skulls excluding the lower jaw) from X-ray images \citep{Bookstein1991} (see Figure \ref{fig:rat-skulls}).

The goal of this regression is to estimate the shape of a rat skull of a 4.5 day old rat, based on 18 rat skulls of ages 7, 14, 21, 30, 40, 60, 90, and 150 days. More formally, our task is  to estimate parameters $\theta\in \Sigma(k, d),\, \phi\in T_{\theta}\Sigma(k, d)$ such that the difference between the geodesic value $f_{\theta}(X)$ for each input $X$ and the given value in the data set $y$ is minimal (with respect to the Riemannian distance function). Here, we set
\begin{align}\label{f-theta}
    f_{\theta}(X) = \exp_{\theta}(X\phi )\in \Sigma(k, d),
\end{align}
where $X$ is the time in days. Once we know the parameters $\theta, \phi$, we can then estimate the shape of a rat's skull for different times $X$ by just plugging in $X$ into $f_{\theta}.$  Below, we show how this task can be done conveniently using the implementation of landmark sets from the \texttt{shape} module and the learning algorithm from Geomstats. Figure \ref{fig:rat-skulls} shows the plot of the resulting estimate. The Kendall shape space \texttt{kendall\_space} can be created as above, by specifying 8 landmarks.

\begin{minted}{python}
import geomstats.backend as gs
from geomstats.learning.geodesic_regression import GeodesicRegression

# Instantiate estimator
gr = GeodesicRegression(
    space=kendall_shape,
    center_X=False,
    method="riemannian",
    initialization="warm_start")

# Fit regression: X: times; y: array of skulls
gr.fit(X, y)

estimated_skull = gr.predict(gs.array([4.5]))
\end{minted}

\subsubsection{Opportunities and challenges when computing with landmarks sets} Representing objects as sets of landmarks, and object shapes as elements of a Kendall shape space, is appealing for several reasons. First, this representation is compact, in the sense that it has a very low memory footprint. In other words, only a few landmarks can distill the essence of a shape. This leads to fast computations and machine learning algorithms that typically require smaller datasets. Second, the differential geometry of the Kendall shape space has been extensively studied \cite{Kendall1993, Dryden2016}. Consequently, it is possible to derive important insights on the behavior of statistical and machine learning algorithms through analytical computations on this shape space. For example, researchers have leveraged Riemannian geometry to study the statistical properties of the mean shape in spaces of landmark shapes including Kendall shape spaces. In particular, it has been observed that for shape spaces without a quotient by the action of scaling, there is an asymptotic bias on the computation of the mean shape. Even with an infinite number of data points, the true mean shape cannot be recovered exactly such that researchers need to resort to bias correction techniques~\cite{Miolane2017TemplateBias}.

Yet, describing objects as landmark sets and their shapes has some drawbacks. First, it requires that the landmarks are correctly positioned next to points of semantic interest: in our example above, next to the actual location of anatomical landmarks on a rat's skull shape. Next, beyond the example of triangles in 2D and triangles in 3D, Kendall shape spaces might be harder to leverage for interpretation purposes.

\subsection{Curves}

\subsubsection{Spaces of curves}

\begin{figure*}[h]
  \centering
  \includegraphics[width=\textwidth]{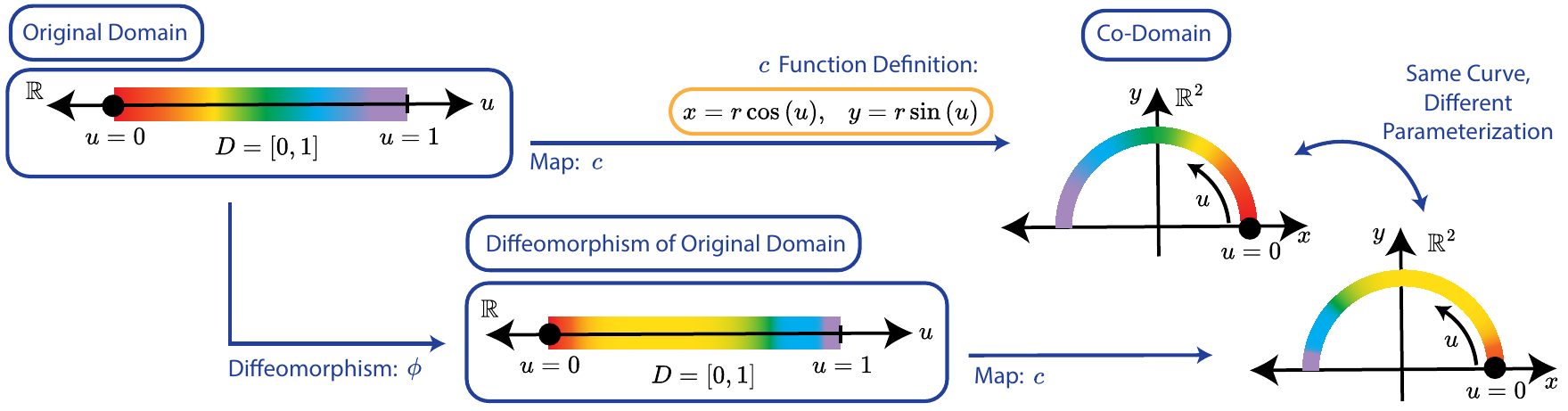}
    \caption{\textbf{Curves.} Consider a planar curve  represented by a function $c: [0, 1] \to \mathbb{R}^2$ which maps a parameter $u\in [0,1]$ to points in $\mathbb{R}^2$: for example, $c$ is a half-circle (top row). The curve's parametrization can be changed by applying a diffeomorphism $\phi$ to the domain $D=[0,1]$ before mapping to $\mathbb{R}^2$ (bottom row). This reparametrization does not change the shape of the curve, which is still a half circle.}
    \label{fig:continuous_curve_definition}
\end{figure*}

Consider the (infinite-dimensional) manifold $\operatorname{Imm}\left(D, \mathbb{R}^d\right)$, whose points are immersions with domain $D$ either the interval $I = [0, 1]$ or the circle $S^1$ \citep{bauer2024elastic}:
$$
\operatorname{Imm}\left(D, \mathbb{R}^d\right)=\left\{c \in C^{\infty}\left(D, \mathbb{R}^d\right): c^{\prime}(u) \neq 0\  \forall u \in D\right\}.
$$
Each immersion $c$ represents a regular smooth parameterized curve in a Euclidean space $\mathbb{R}^d$, denoted $c: D \rightarrow \mathbb{R}^d$. The curve $c$ is open if $D$ is the interval $[0, 1]$; closed if $D$ is the circle $S^1$. An example of open curve in $\mathbb{R}^2$ is shown in Figure~\ref{fig:continuous_curve_definition}.

In shape analysis, one is not interested in the parametrized curve itself, but only in its features after quotienting out the action of shape-preserving transformations: translations, rotations and reparametrizations. To deal with translations, we restrict to the space of parameterized curves starting at the origin in Euclidean space
\begin{equation}
     \operatorname{Imm}_0\left(D, \mathbb{R}^d\right) = \operatorname{Imm}\left(D, \mathbb{R}^d\right) / \mathbb{R}^d.
\end{equation}
Just like for landmarks, the group of rotations $SO(d)$ acts by left multiplication on this space of curves starting at the origin. As for reparametrizations, they are represented by elements of an appropriate group of diffeomorphisms $\mathcal{D}(D)$
and act by right composition on $\operatorname{Imm}_0\left(D, \mathbb{R}^d\right)$:
$$
\operatorname{Imm}_0\left(D, M\right) \times \mathcal{D}(D) \rightarrow \operatorname{Imm}_0\left(D, \mathbb{R}^d\right), \quad(c, \varphi) \mapsto c \circ \varphi.
$$
This action merely changes the parametrization of the curve but not its actual shape, as shown in Figure~\ref{fig:continuous_curve_definition}.

We then define a Riemannian metric on $\operatorname{Imm}_0\left(D, \mathbb{R}^d\right)$ that is invariant with respect to these shape preserving transformations. A popular family of metrics with this property is given by the elastic metrics \citep{mio2007shape,bauer2024elastic}, defined for scalars $a, b$ as
\begin{equation}\label{eq:elastic-bilinear-form}
    G_c^{a, b}(h, k)=\int_D a^2\langle D_s h^{\perp}, D_s k^{\perp}\rangle+b^2\langle D_s h^{\top}, D_s k^{\top}\rangle d s.
\end{equation}
Here, 
$h, k$ are tangent vectors to the manifold of curves 
at a point $c$, \textit{i.e.}, vector fields along $c$ that represent infinitesimal deformations of the curve; 
$D_s=\frac{1}{\left|c^{\prime}\right|} \frac{d}{d u}$, $d s=\left|c^{\prime}\right| d u$ represent differentiation and integration with respect to the arc length parameter $s$; 
$\bullet^{\top}$ and $\bullet^{\perp}$ denote projection onto the normal and tangential parts of a tangent vector, \textit{i.e.},
$
D_s h^{\top}=\left\langle D_s h, \frac{c^{\prime}}{\left|c^{\prime}\right|}\right\rangle \frac{c^{\prime}}{\left|c^{\prime}\right|},
$
where $\langle \cdot, \cdot \rangle$ is the Euclidean inner product. The coefficients $a$ and $b$ in the $G^{a,b}$ metric indicate the extent to which bending and stretching the curve are respectively penalized.



Besides their reparametrization-invariant property and interpretation from linear elasticity theory (see \citep{bauer2024elastic}), elastic metrics are appealing due to the existence of closed-form solutions for geodesics. For example, \citep{bauer2024elastic}  shows that the square root velocity (SRV) transform $R$ \citep{Srivastava2011} defined as:
\begin{equation} \label{eq:srv}
R: \operatorname{Imm}_0\left([0,1], \mathbb{R}^d\right) \rightarrow C^{\infty}\left([0,1], \mathbb{R}^d \backslash\{0\}\right)
\end{equation}
$$
c \mapsto \frac{c^{\prime}}{\sqrt{||c^{\prime}||}},
$$
is an isometry between the manifold $\operatorname{Imm}_0([0,1], \mathbb{R}^d)$ of Euclidean curves starting at the origin equipped with the elastic metric $G^{a, b}$,
and the space $C^{\infty}([0,1], \mathbb{R}^d \backslash\{0\})$ equipped with a multiple $4 b^2 G^{L_\lambda^2}$ of a conic Riemannian metric defined in~\citep{bauer2024elastic}. Geodesics are known in closed form for the metric $G_\lambda^2$, which allows us to compute geodesics in the original space of curves $\operatorname{Imm}_0\left([0,1], \mathbb{R}^d\right)$. Indeed, we convert an initial point and initial tangent vector in $\operatorname{Imm}_0\left([0,1], \mathbb{R}^d\right)$ into a point and tangent vector in  $C^{\infty}\left([0,1], \mathbb{R}^d \backslash\{0\}\right)$ using the transform $R$ and its differential, compute the geodesic there, and use the inverse transform $R^{-1}$ to obtain the corresponding geodesic in $\operatorname{Imm}_0\left([0,1], \mathbb{R}^d\right)$. 
In particular, when $a=1, b=1/2$, $G^{L_\lambda^2}$ reduces to a standard $L^2$ metric on $C^{\infty}\left(I, \mathbb{R}^d \backslash\{0\}\right)$, \textit{i.e.}, a metric resulting from endowing the space with the $L^2$-inner product. 
The metric $G^{1, 1/2}$ is known as the square root velocity (SRV) metric~\citep{Srivastava2011}.

\subsubsection{Spaces of curve shapes}

Recall that we have defined shapes to be what is left after factoring out certain transformations. Depending on the application, we can thus define the space of curve shapes to be the quotient of $\operatorname{Imm}_0\left(D, \mathbb{R}^d\right)$ under the action of either reparametrizations $\mathcal{D}(D)$, rotations $\operatorname{SO}(d)$ or both $\mathcal{D}(D) \times \operatorname{SO}(d))$:
\begin{align*}
    \mathcal{Q}_{\text{reparam.}} 
    &= 
    \operatorname{Imm}_0\left(D, \mathbb{R}^d\right) / \mathcal{D}(D) \\
    \mathcal{Q}_{\text{rotations}} 
    &= 
    \operatorname{Imm}_0\left(D, \mathbb{R}^d\right) / \operatorname{SO}(d)\\
    \mathcal{Q}
    &=
    \operatorname{Imm}_0\left(D, \mathbb{R}^d\right) / ( \mathcal{D}(D) \times \operatorname{SO}(d)).
\end{align*}
The elastic metric $G^{a, b}$ on $\operatorname{Imm}_0\left(D, \mathbb{R}^d\right)$ is invariant under both actions, \textit{i.e.}:
$$G^{a, b}_c(h, k)=G^{a,b}_{R(c \circ \varphi)}(R(h \circ \varphi), R(k \circ \varphi)) \quad \forall\phi \in \mathcal{D}(D), \,R\in \operatorname{SO}(d),$$
and induces a Riemannian metric and a quotient distance~\eqref{eq:quotient-dist} on all possible quotient spaces, as described in Section~\ref{sec:quotient_dist}.

\subsubsection{Aligning curves}

Algorithms to align curves are central to perform computations on the respective quotient spaces. 

\textit{Aligning curves in rotations.} As with landmark sets, the optimal rotation that aligns one curve onto another can be found by a Procrustes approach, specifically through the singular value decomposition of a quantity involving only the 
SRV representations of the two curves.

\textit{Aligning curves in reparametrizations.} Alignment is particularly challenging when quotienting out reparametrizations due to the infinite-dimensional nature of the reparametrization group. 
The alignment strategies currently implemented in the \texttt{shape} module to find optimal reparametrizations include a dynamic programming-based algorithm \citep{Srivastava2011}, and an iterative horizontal alignment algorithm \citep{lebrigant2019discrete} ~| which we recall below.

The dynamic programming-based algorithm \citep{Srivastava2011} searches for a monotonically increasing diffeomorphism $\gamma$ in a uniform $n \times n$-grid that, for the SRV metric, minimizes the integrand
$$
q_1(t)\cdot q_2(\gamma(t)) \dot{\gamma}(t)^\frac{1}{2}
$$
over the domain, where $q_i$ is the SRV transform of $c_i$. The computation of the integral is done recursively. Assuming the integral is zero at $(0, 0)$, it sequentially computes its value for each point of the grid. 
To limit the computational cost, a numerical parameter $s$ is usually used to limit the degree of looking-back. Specifically, the solution at node $(i, j)$ is found by considering only the subgrid $(i-s, j-s)\times (i-1, j-1)$ instead of the full grid (such parameter restricts the admissible slopes).

The iterative horizontal alignment algorithm \citep{lebrigant2019discrete} finds an horizontal geodesic by successively iterating between two steps: i) construct the geodesic between two curves $c_1, c_2$; ii) compute the horizontal part of the geodesic. In other words, it successively finds a better representative of the curve $c_2$ with respect to $c_1$, \textit{i.e.}, a curve in the fiber of $c_2$ that is closer (with respect to the total space metric) to $c_1$. The key ingredient for this procedure is the computation of the horizontal part of the geodesic, which requires solving a linear system (see \citep{lebrigant2019discrete} for details). This algorithm is agnostic to the metric on the total space.

\textit{Aligning curves in rotations and reparametrizations.} Finding the optimal pair of rotation and reparametrization is approximated by a two-step iterative process: i) find the optimal rotation while fixing the parametrization; ii) find the optimal reparametrization while fixing the rotation\cite{srivastava2016book}.

\subsubsection{Curves and their shapes in the \texttt{shape} module}

In the \textit{shape} module, we represent spaces of continuous curves and their shapes by discrete curves with $k$ sampling points, as done in \citep{lebrigant2019discrete}. All the quantities of interest can be obtained from this discrete curves representation by standard numerical methods, such as finite differences and numerical integration. We note that points are considered to be uniformly sampled with respect to their parameterization.

The (finite-dimensional) discrete curves space with $k$ sampling points is defined as a product manifold $M = \mathbb{R}^d \times \cdots \times \mathbb{R}^d$. 
It is implemented in the Python class \codeobj{DiscreteCurves}, which inherits from \codeobj{NFoldManifold}. By default, it is equipped with the standard discrete $L^2$ metric, implemented in the Python class \codeobj{L2CurvesMetric}, which inherits from the class \codeobj{NFoldMetric}. For example, a space of discrete curves in 2D with 200 sampling points can be instantiated with the following code:

\begin{minted}{python}
from geomstats.geometry.discrete_curves import DiscreteCurves

curves = DiscreteCurves(k_sampling_points=200, ambient_dim=2, starting_at_origin=False)
\end{minted}


Removing translations is achieved 
by considering curves that start at the origin. In practice, we subtract the initial point and omit it from the representation. 


\begin{minted}{python}
curves = DiscreteCurves(k_sampling_points=200, ambient_dim=2, starting_at_origin=True)
\end{minted}

This space can be equipped with the elastic metric $G^{a, b}$ and in particular the SRV metric $G^{1, \frac{1}{2}}$, respectively implemented in the classes \codeobj{ElasticMetric} and \codeobj{SRVMetric}. Taking advantage of the isometry \eqref{eq:srv}, we have implemented these metrics as children of the class \codeobj{PullbackDiffeoMetric}, which implements the general structure of pullback metrics by diffeomorphisms in Geomstats.
Similarly, the operations related to the SRV 
transform \eqref{eq:srv} are encoded in the Python class \codeobj{SRVTransform} 
which inherits from the abstract class \codeobj{Diffeo} that implements the structure of diffeomorphisms in Geomstats. 

\begin{figure}[h]
\centering
\includegraphics[width=0.9\linewidth]{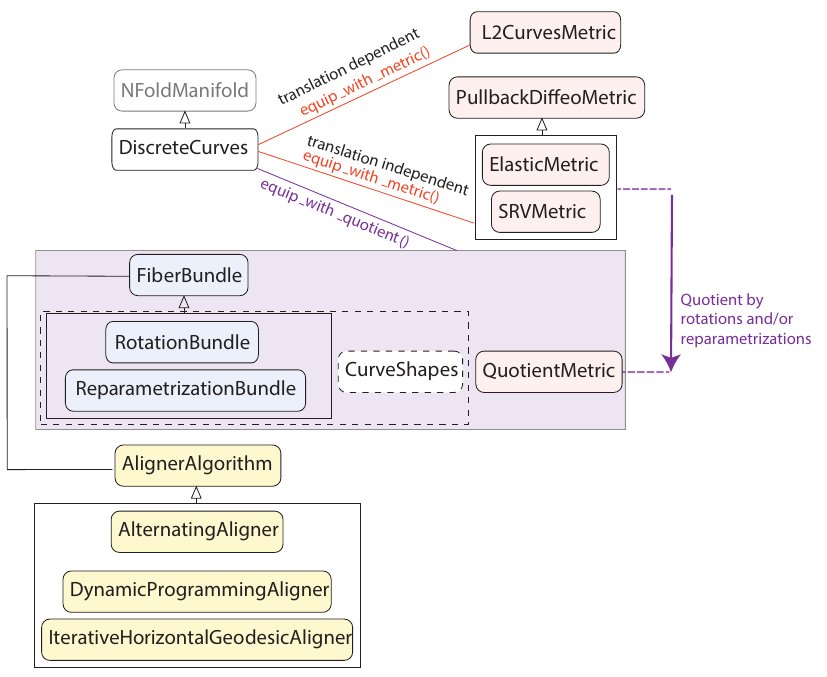}
\caption{\textbf{Curves and their shapes in the \texttt{shape} module of Geomstats.} The Python class \texttt{DiscreteCurves} (left) represents a space of curves, a space of curves that start at the origin, or a space of curve shapes where either rotation or reparametrization or both have been quotiented, depending on the Riemannian metric that equips it: with the $L_2$ Euclidean metric, the elastic or the Square-Root-Velocity (SRV) metric, or a quotient elastic or SRV metric where the quotient is shown by the purple arrow, respectively (right). Each of these metrics quotients specific group actions: translations, rotations and/or reparametrizations. We note that, while the elastic and SRV metrics do not depend on translations, they are \textit{not} built as quotients of the translation-dependent $L_2$ Euclidean metric. We indicate the abstract Python classes from which each class inherits, with empty black arrows representing inheritance. The alignment algorithms that perform quotients by rotations and reparametrizations are indicated in yellow.}
\label{fig:curves-structure}
\end{figure}


Next, in order to create a space of curve shapes, we equip the space of curves with the action of the group of rotations $SO(d)$, and/or of the group of reparametrizations, which we quotient out. The quotient space is the space of curve shapes.

\begin{minted}{python}
discrete_curves.equip_with_group_action(("rotations", "reparametrizations"))
discrete_curves.equip_with_quotient()

curve_shapes = discrete_curves.quotient
curve_shapes_metric = curve_shapes.metric
\end{minted}

Recall that when a manifold is equipped with a metric and a group action, the method \codeobj{equip\_with\_quotient()} does two things.

First, it creates a quotient space equipped with a quotient metric. Therefore, the shape space comes equipped with a Riemannian metric, which is the one induced by the metric on the space of curves; by default, these are the SRV metric and the quotient SRV metric. The user can access essential computations of Riemannian geometry, such as \texttt{curve\_shapes\_metric.exp()}, \texttt{curve\_shapes\_metric.log()} and \texttt{curve\_shapes\_metric.geodesic()}.

Second, it equips the total space with a fiber bundle. Two children of the class \codeobj{FiberBundle} are implemented: \codeobj{ReparametrizationBundle} and \codeobj{RotationBundle}. The reparametrization alignment algorithms \codeobj{DynamicProgrammingAligner} and \codeobj{IterativeHorizontalGeodesicAligner} are implemented as children of \codeobj{AlignerAlgorithm}, and can be passed by composition to an instance of \codeobj{ReparametrizationBundle}. The general structure and code provided by the Python class \codeobj{QuotientMetric} performs the rest of the computations.

\subsubsection{Use case: Statistics on curve shapes}

\begin{figure}[ht]
\centering
\includegraphics[width=0.9\textwidth]{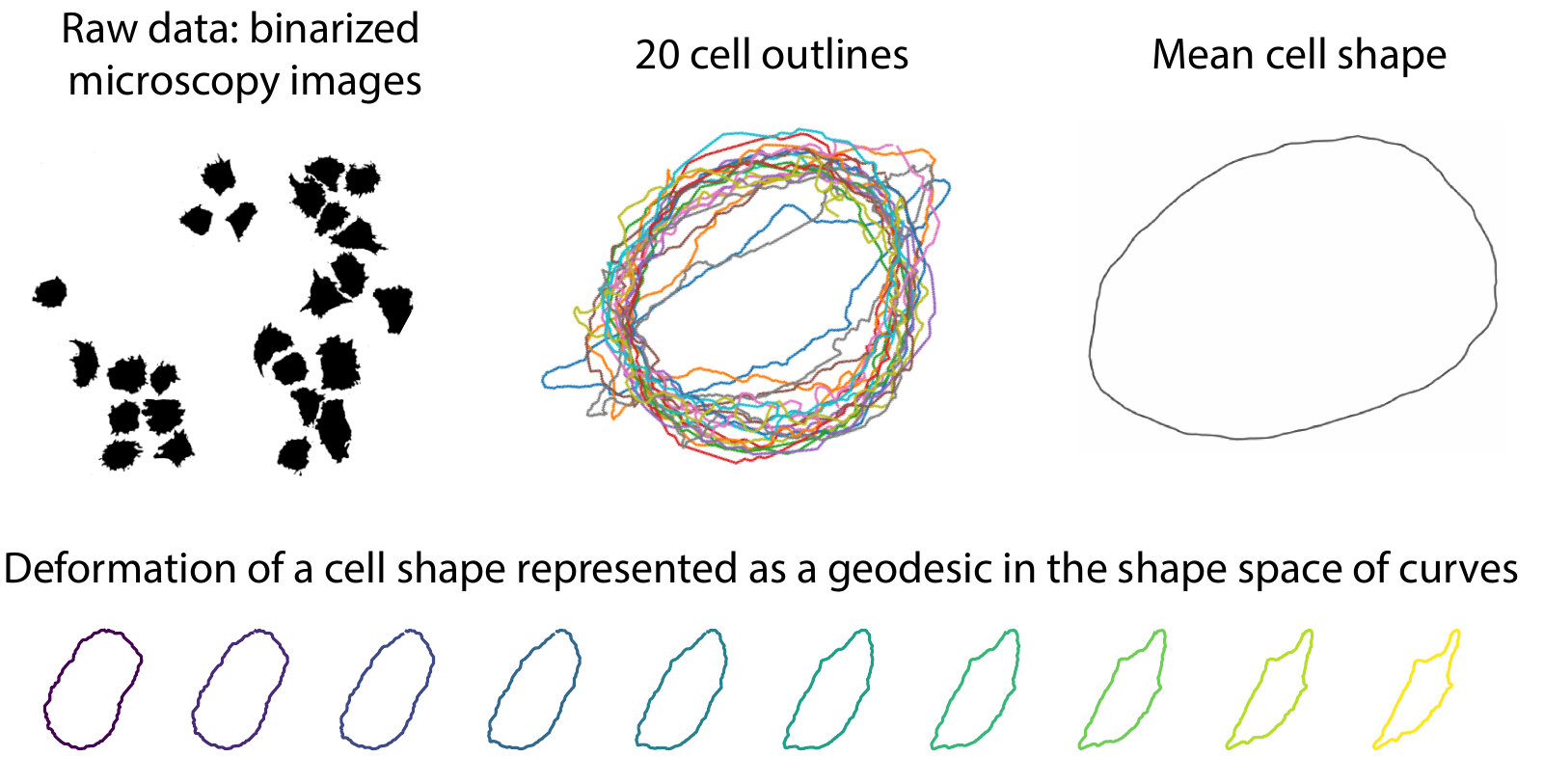}
    \caption{\textbf{Statistics on cancer cell shapes with the \texttt{shape} module of Geomstats.} Top-left: Binarized image obtained from microscopy and containing a set of cancer cells. Top-middle: 20 cells from the data set. Top-right: Frech\'et mean of the 20 cells shown on the left. Bottom: Geodesic between two cancer cells, going from a less aggressive cancer to a more aggressive cancer, showing the corresponding characteristic deformation of the cell outline.}
    \label{fig:combined}
\end{figure}

We show how to perform computations and calculate 
summary 
statistics on curve shapes implemented in the \texttt{shape} module using algorithms available in the main Geomstats package. Specifically, we study a data set of mouse \textit{Osteosarcoma} imaged cells (AXCFP2019), see Figure~\ref{fig:combined}. Each curve is represented by 200 uniformly-spaced sampling points. To showcase how cells can be compared in Geomstats, we start by showing how to compute the geodesic between two cells with respect to the SRV metric (equipped by default), see Figure~\ref{fig:combined} (D). 

\begin{minted}{python}
from geomstats.geometry.discrete_curves import DiscreteCurves

# Instantiate space
space = DiscreteCurves(k_sampling_points=200, ambient_dim=2, starting_at_origin=True)

# Compute geodesic and evaluate it at midpoint
geod_func = space.metric.geodesic(initial_point=cell_start, end_point=cell_end)
geod_func(0.5)
\end{minted}

The mean of 20 of such cells is showed in Figure~\ref{fig:combined} (C). The following code snippet shows how it can be computed in Geomstats.

\begin{minted}{python}
from geomstats.learning.frechet_mean import FrechetMean

# Instantiate and fit estimator
mean = FrechetMean(space)
mean.fit(cell_shapes)

# Access mean shape
mean_estimate = mean.estimate_
\end{minted}

\subsubsection{Opportunities and challenges when computing with curves}
Computing with curves and their shapes is appealing for objects that do not have well-defined, semantic landmarks, but rather contours. For example, a cell border typically does not have landmarks, yet it has a well-defined contour. The differential geometry theory of spaces of curves and curve spaces is well-defined and relatively mature. However, and perhaps surprisingly, there exist several outstanding challenges when it comes to their numerical implementations. First, while the theory for closed curves is well-defined, as a submanifold of the space of open curves, the implementation of closed curves is often ad-hoc, 
resorting to 
various techniques to close the curves. Second, the algorithms performing alignment in reparametrization can be inaccurate and slow. Third, while the theory prescribes to quotient the joint action of rotations and reparametrization, in practice researchers quotient each sequentially. This approach seems justified by the fact that the results are often only minimally affected. Lastly, while the theory proposes elastic Riemannian metric on curve spaces for any real scalar $a, b$, we observe in practice that larger or finer values of $a,b$ or their ratios can deteriorate the quality of the numerical results, making the choice of the parameters particularly challenging.

Meanwhile, this produces exciting research directions. 
An open research direction is to investigate whether changing parameters $a, b$ in computations and machine learning algorithms can lead to different, interpretable results upon extracting knowledge from curves in natural sciences. A related research direction also performs Riemannian metric learning, \textit{i.e.}, learns the parameters $a, b$ that best describe a trajectory of curve shapes as a geodesic~\cite{myers2022regression}.

\subsection{Surfaces}

\subsubsection{Spaces of surfaces}

\begin{figure*}[h!]
  \centering
  \includegraphics[width=\textwidth]{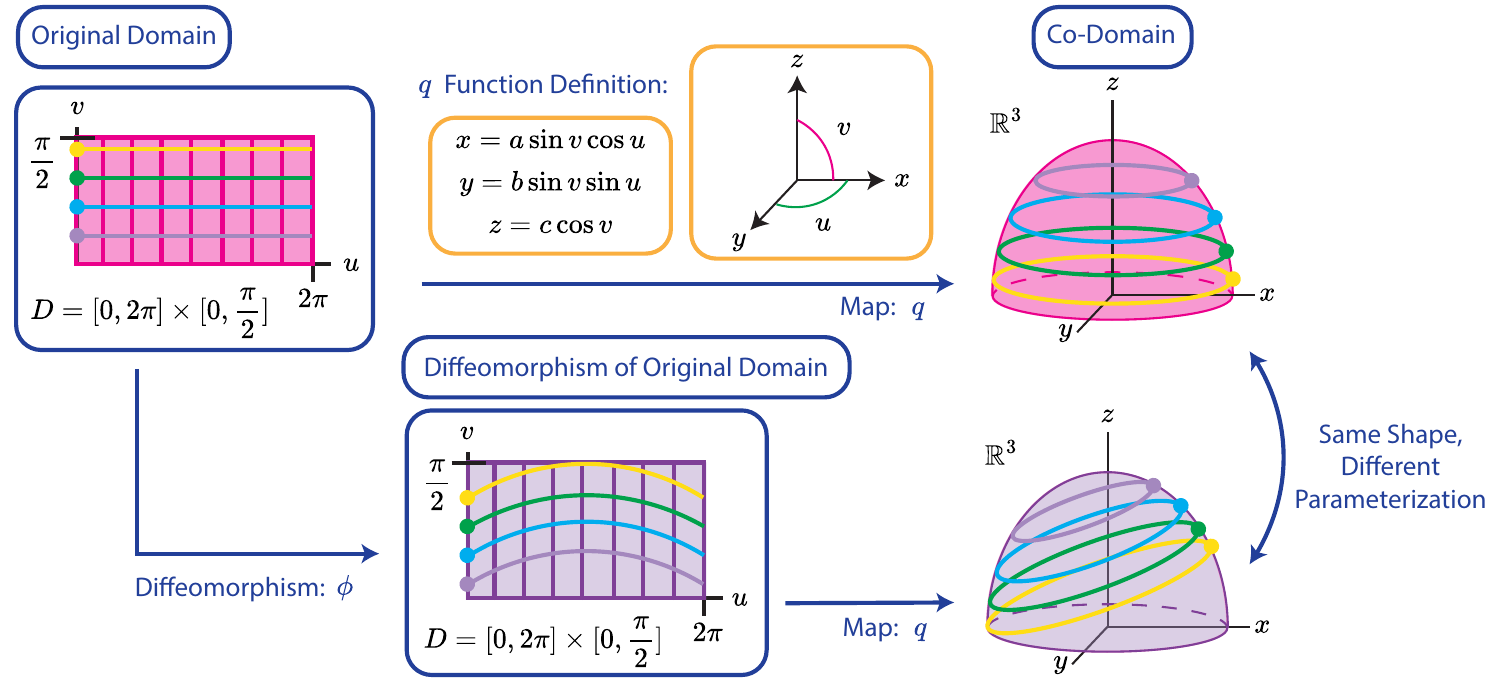}
    \caption{\textbf{Surfaces}. Consider a surface represented by a function $q: D \to \mathbb{R}^3$ that maps parameters $(u, v)\in D$ to points in 3D space $q(u, v) \in \mathbb{R}^3$, forming a half sphere (top row). The surface's parameterization can be changed by applying a diffeomorphism $\phi$ to the domain $D$ before mapping to $\mathbb{R}^3$ (bottom row). This reparametrization does not change the shape of the surface, which is still a half sphere. Figure and caption adapted from \cite{myers2023geodesic}.}
    \label{fig:continuous_surface_definition}
\end{figure*}

Spaces of surfaces are defined analogously to spaces of curves introduced in the previous subsection. The space of regular smooth parameterized immersed surfaces in $\mathbb{R}^3$ can be identified with the (infinite-dimensional) manifold:
\begin{equation}
    \operatorname{Imm}\left(D, \mathbb{R}^3\right)=\left\{q \in C^{\infty}\left(D, \mathbb{R}^3\right): q^{\prime}(u) \neq 0\  \forall u \in D\right\},
\end{equation}
where $D$ is a compact 2-dimensional space of parameters, for example $D=[0, 2\pi] \times [0, \frac{\pi}{2}]$. This is shown in Fig.~\ref{fig:continuous_surface_definition}, adapted from \cite{myers2023geodesic}.

A second-order family of Sobolev metrics can be defined on $\operatorname{Imm}(D, \mathbb{R}^3)$ \citep{hartman2023elastic}:
\begin{equation} \label{eq:surfaces-metric}
\begin{aligned}
G_q(h, k) & =\int_D a_0\langle h, k\rangle+a_1 g_q^{-1}\left(d h_m, d k_m\right) \\
& +b_1 g_q^{-1}\left(d h_{+}, d k_{+}\right)+c_1 g_q^{-1}\left(d h_{\perp}, d k_{\perp}\right) \\
& +d_1 g_q^{-1}\left(d h_0, d k_0\right) + a_2 \langle\Delta_q h, \Delta_q h\rangle  \operatorname{vol}_q,
\end{aligned}
\end{equation}
where $q \in \operatorname{Imm}(D, \mathbb{R}^3)$, $h, k \in C^{\infty}\left(D, \mathbb{R}^3\right)$ are tangent vectors to the manifold of surfaces. In practice, $h$ is a vector field along the surface, which has tangential and normal components with respect to the surface. The differential $dh$ can be interpreted as a vector-valued one form (\textit{i.e.}, a map from the tangent bundle $T D$ to $\mathbb{R}^3$) and $g_q$ refers to the Riemannian metric \textit{of the surface $q$}, \textit{i.e.}, the pullback of the Euclidean metric of the 3D ambient space onto the 2D surface defined by $q$~\citep{hartman2023elastic}. Lastly, $\operatorname{vol}_q$ is the surface area measure of the smooth immersion $q$. Each choice of parameters $(a_0, a_1, b_1, c_1, d_1, a_2)$ gives a different Riemannian metric from the family of metrics.

This metric results from the weighted combination of a zeroth-order term:
$$
G_q^{0\text{th}}(h, k)=\int_D\langle h, k\rangle \operatorname{vol}_q,
$$
with a first-order term:
$$
G_q^{1\text{st}}(h, k) = \int_D g_q^{-1}(d h, d k) \operatorname{vol}_q,
$$
and a second-order term:
$$
G^{2\text{nd}}_q(h, k) = \int_D \left\langle\Delta_q h, \Delta_q h\right\rangle \operatorname{vol}_q.
$$

The zeroth-order term $G_q^{0\text{th}}(h, k)$ is the $L^2$ metric on $q \in \operatorname{Imm}(D, \mathbb{R}^3)$, which is degenerate \citep{hartman2023elastic}. The first-order term $G_q^{1\text{st}}(h, k)$ is introduced to avoid this degeneracy and can be expanded, as shown in Equation~\eqref{eq:surfaces-metric}, by the decomposition of $dh$ into physically-interpretable components \cite{su2020surfaces, hartman2023elastic}: a shearing term $d h_m$, a scaling term $d h_+$, and a bending term $d h_{\perp}$, and a less interpretable $d h_0$ term. This physical interpretation is the reason first-order Sobolev metrics on $\operatorname{Imm}(D, \mathbb{R}^3)$ are known as elastic metrics. The second-order term $G^{2\text{nd}}_q(h, k)$ is introduced due to empirical evidence that first-order terms are still too weak to prevent geodesics from leaving the space of immersions which leads to instability in numerical schemes \citep{hartman2023elastic} and involves the computation of the Laplacian $\Delta_q$.

\subsubsection{Spaces of surface shapes}

We recall that second-order Sobolev metrics $G_q(h, k)$ on $\operatorname{Imm}(D, \mathbb{R}^3)$ are invariant under the action of the reparametrization $\mathcal{D}(D)$, rotation $\operatorname{SO}(3)$, and translation $\mathbb{R}^3$ groups \citep{hartman2023elastic}:
$
G_q(h, k)=G_{R(q \circ \varphi)+\tau}(R(h \circ \varphi), R(k \circ \varphi))
$
where $\phi \in \mathcal{D}(D)$, $R\in \operatorname{SO}(3)$, and $\tau \in \mathbb{R}^3$. Thus, we can define spaces of surface shapes by considering quotient structures on $\operatorname{Imm}(D, \mathbb{R}^3)$ coming from any of those invariances. Here, we are interested on the quotient of reparametrizations $\mathcal{D}(D)$:
\begin{equation*}
    \mathcal{Q}_{\text{reparam.}} 
    = 
    \operatorname{Imm}(D, \mathbb{R}^3) / \mathcal{D}(D).
\end{equation*}

The reparametrization-invariance of $G_q(h, k)$ induces a quotient metric on $\mathcal{Q}_{\text{reparam.}}$ the spaces of surface shapes in $\mathbb{R}^3$.

\subsubsection{Aligning surfaces in reparametrization}

As with discrete curves, the action of the (infinite-dimensional) reparametrization group on discrete surfaces is particularly challenging to implement. Following \citep{hartman2023elastic}, we perform alignment by solving the geodesic boundary value problem in the quotient space of unparametrized discrete surfaces. Specifically, we follow their relaxed optimization formulation which uses a discrepancy term given by kernel metrics on (oriented) varifold representations of surfaces (see \citep{kaltenmark2017varifold} for details on varifolds).



\subsubsection{Surfaces and their shapes in the \texttt{shape} module} \label{sec:discrete-setting-surfaces}

\begin{figure*}[h]
    \centering
    \includegraphics[width=\textwidth]{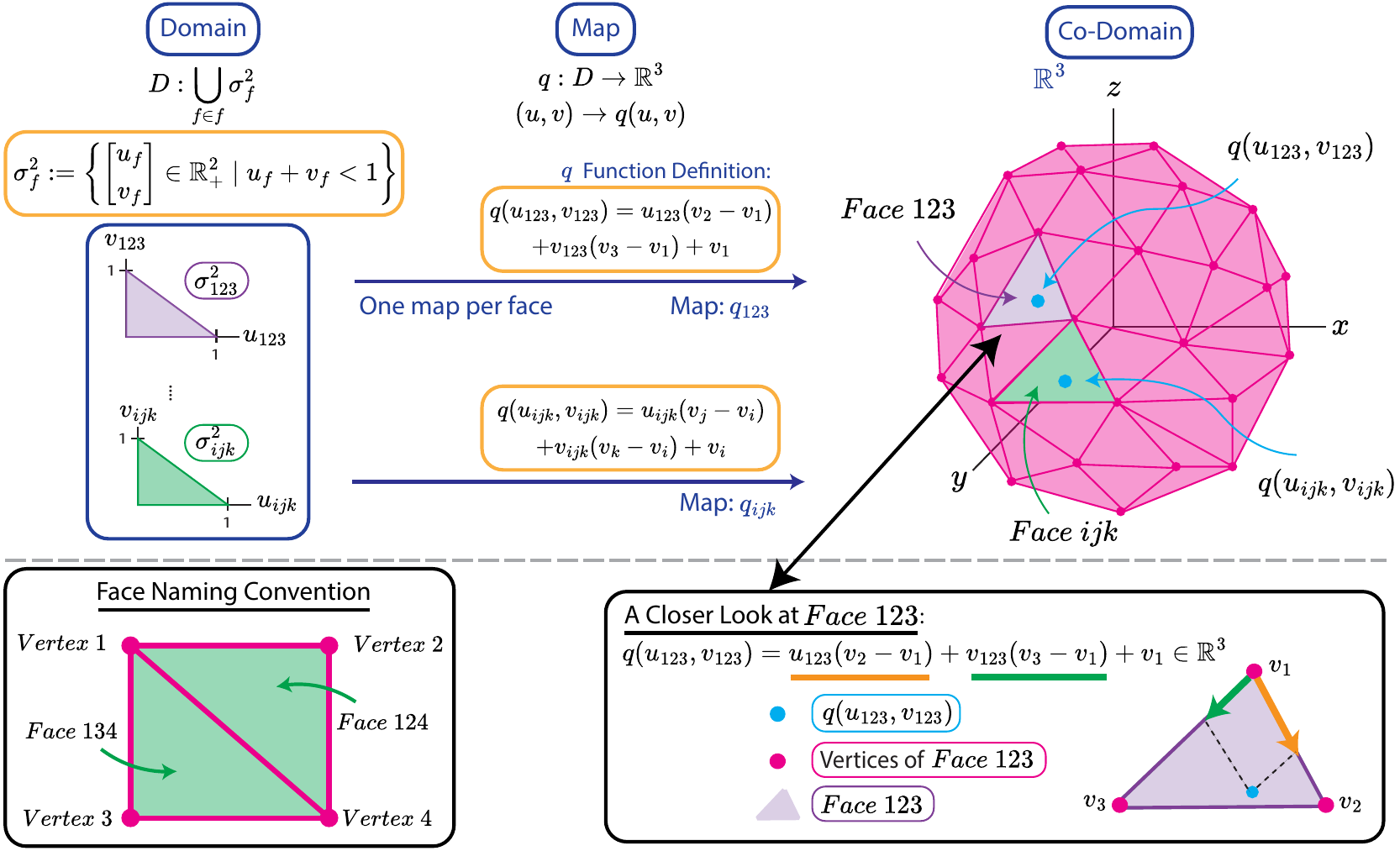}
    \caption{\textbf{Discrete Surfaces.} A discrete surface can be represented by a map $q: D \to \mathbb{R}^3$. Each ``sub-domain'' in $D$ corresponds to a different face on the discrete surface in $\mathbb{R}^3$. We can change the parameterization of the discrete surface by changing the number of sub-domains in $D$, or by changing the vertices corresponding to one or more of the sub-domains.}
    \label{fig:discrete_surface_definition}
\end{figure*}

In the \textit{shape} module, we follow \citep{hartman2023elastic} and represent surfaces by triangle meshes with fixed combinatorial structure, \textit{i.e.}, meshes with fixed number of vertices $V$ and connectivity (set of edges $E$ and faces $F$). We call these meshes the discrete surfaces. Let $\mathfrak{M}$ represent the set of such triangle meshes. As shown in Fig.~\ref{fig:discrete_surface_definition}, a mesh $q \in \mathfrak{M}$ is fully defined by the position of its vertices and can be seen as a piecewise linear surface.

\begin{figure}[h]
\centering
\includegraphics[width=0.9\linewidth]{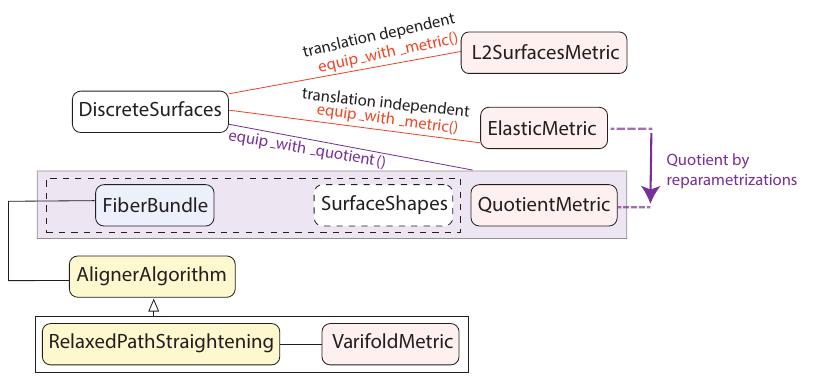}
\caption{\textbf{Surfaces and their shapes in the \texttt{shape} module of Geomstats.} The Python class \texttt{DiscreteSurfaces} (left) represents a space of surfaces, a space of surfaces whose barycenter is located at the origin, or a space of surface shapes where reparametrization has been quotiented, depending on the Riemannian metric that equips it: with the $L_2$ Euclidean metric, the elastic metric, or a quotient elastic metric where the quotient is shown by the purple arrow, respectively (right). Each of these metrics quotients specific group actions: translations and reparametrizations. We note that, while the elastic metric does not depend on translations, it is \textit{not} built as quotient of the translation-dependent $L_2$ Euclidean metric.  We indicate the abstract Python classes from which each class inherits, with empty black arrows representing inheritance. The alignment algorithm that performs quotient by reparametrizations is indicated in yellow.}
\label{fig:surfaces-structure}
\end{figure}

In the \texttt{shape} module, the Python class \codeobj{DiscreteSurfaces} inherits from the \codeobj{Manifold} class and implements the space of discrete surfaces (see Figure~\ref{fig:surfaces-structure} for details). A space of surfaces with 4 vertices and two triangular faces can be instantiated as:

\begin{minted}{python}
from geomstats.geometry.discrete_surfaces import DiscreteSurfaces

faces = gs.array([[0, 1, 2], [1, 2, 3]])
discrete_surfaces = DiscreteSurfaces(faces=faces)
\end{minted}

In order to use Riemannian geometry to compute with discrete surfaces, the discrete counterparts of several smooth quantities, including the Riemannian metric, need to be defined \citep{hartman2023elastic}. In the field of discrete differential geometry \citep{crane2018discrete}, we may discretize tangent vectors on the vertices $V$, first-order terms on the faces $F$, the Laplacian on the dual cell \citep{botsch2010polygon}, and the surface area measure both at vertices and at faces. With these discretization choices, the discrete counterpart of the Sobolev metric of Equation~\eqref{eq:surfaces-metric} is found in \citep{hartman2023elastic}:
$$
\begin{aligned}
& G_q(h, k)=\sum_{v \in V} a_0\langle h, k\rangle \operatorname{vol}_v \\
& \quad+\sum_{f \in F}\left(a_1 g_f^{-1}\left(d h_m, d k_m\right)+b_1 g_f^{-1}\left(d h_{+}, d k_{+}\right)\right. \\
& \left.\quad+c_1 g_f^{-1}\left(d h_{\perp}, d k_{\perp}\right)+d_1 g_f^{-1}\left(d h_0, d k_0\right)\right) \operatorname{vol}_f \\
& \quad+\sum_{v \in V} a_2\left\langle\Delta_q h, \Delta_q k\right\rangle \operatorname{vol}_v.
\end{aligned}
$$

We refer the reader to \citep{hartman2023baresa} for detailed expressions for each term. These computations are implemented in the \codeobj{ElasticMetric} class, which inherits from the \codeobj{RiemannianMetric} class. The class \codeobj{DiscreteSurfaces} and can be equipped with an elastic metric object.

Next, we equip the space of surfaces with the action of the group of reparametrizations, which we quotient out. The quotient space is the space of surface shapes.

\begin{minted}{python}

discrete_surfaces.equip_with_group_action("reparametrizations")
discrete_surfaces.equip_with_quotient()

surface_shapes = discrete_surfaces.quotient
surface_shapes_metric = surface_shapes.metric
\end{minted}

The shape space comes equipped with a Riemannian metric, which by default is the quotient elastic metric. The user can access essential computations of Riemannian geometry, such as \texttt{surface\_shapes\_metric.exp()}, \texttt{surface\_shapes\_metric.log()} and \texttt{surface\_shapes\_metric.geodesic()}. The quotient structure relies on the alignment procedure discussed above, where the class \codeobj{VarifoldMetric} implements (oriented) varifold distances within the framework of discrete surfaces \cite{hartman2023elastic} by leveraging the fast kernel operations implementation available in Keops \citep{charlier2021keops}.

\textit{Numerical approaches to compute geodesics.} Given the absence of closed-form solutions for most of the metric-related quantities, \codeobj{ElasticMetric} heavily relies on numerical methods, which we describe here. For example, there are no closed-form expressions for the exponential and logarithm maps, which thus need to be computed numerically from the geodesic equation. The geodesic equation for $(\operatorname{Imm}_0(D, \mathbb{R}^3), G_q(h, k))$ is a non-linear, second-order in time, fourth-order in space, partial differential equation \citep{hartman2023elastic}. To avoid solving it directly, we find geodesics on $(\mathfrak{M}, G_q(h, k))$ through algorithms that rely on numerical optimization, usually limited memory BFGS algorithm \citep{liu1989lbfgs} combined with automatic differentiation for gradient computation \citep{paszke2019pytorch, maclaurin2015autograd}. These methods find the optimal location of the vertices of a path of discrete surfaces, represented as a path of triangle meshes \citep{rumpf2015geodesic}.

First, we present the numerical approach to solve the geodesic boundary value (BVP) problem, \textit{i.e.}, the problem of finding the geodesic in surface space connecting two surfaces. This approach is called \textit{path straightening}. A piecewise-linear path of $N$ uniformly-sampled triangle meshes $q = (q(t_i))_{i \in J}$, for $J = \{0, \cdots, N - 1\}$ and $t_i = i / (N - 1)$, can be ``straightened'' into a geodesic by minimizing the Riemannian energy, given in discrete form by:
\begin{equation} \label{eq:discrete-riem-energy-surfaces}
    E(q) = \frac{1}{2N} \sum_{i=0}^{N-1} G_{q(t_i)}(\dot{q}(t_i), \dot{q}(t_i)).
\end{equation}
Here, $\dot{q}(t_i)$ is obtained by, \textit{e.g.}, forward finite differences:
$
\dot{q}(t_i) = (N-1) (q(t_{i+1}) - q(t_i)).
$
Given a discrete geodesic $q$, the tangent vector $h$ that shoots from $q(t_0)$ to $q(t_{N-1})$ is found through:
$
h = (N - 1) (q(t_1) - q(t_0)).
$
The path-straightening algorithm is implemented via the class \codeobj{PathStraightening}, which is an attribute of the class \texttt{ElasticMetric}.

Second, we present the numerical approach to solve the geodesic initial value (IVP) problem, \textit{i.e.}, the problem of finding the geodesic starting at an initial surface $q(t_0)$ with initial velocity $h$. Here, strategy consists in finding the midpoint $q(t_i)$ of a geodesic segment assuming known endpoints $q(t_{i-1}), q(t_{i+1})$~\citep{rumpf2015geodesic, hartman2023elastic}. This leads to the following numerical optimization problem at each time step \citep{hartman2023elastic}:
\begin{equation} \label{eq:discrete-surfaces-ivp}
q(t_{i+1}) = \argmin _{\bar{q}(t_{i+1})} ||F\left(\bar{q}(t_{i+1}) ; q(t_i), q(t_{i-1})\right)||_2^2,
\end{equation}
where $F\left(q(t_{i+1}) ; q(t_i), q(t_{i-1})\right) = 0$ denotes the following system of equations \citep{hartman2023elastic}:
$$
\begin{aligned}
&2G_{q(t_{i-1})}(q(t_{i}) - q(t_{i - 1}), B_i) - 2G_{q(t_{i})}(q(t_{i + 1})- q(t_{i}), B_i)\\
&+ D_{q(t_{i})} G(q(t_{i+1})-q(t_{i}), q(t_{i+1})-q(t_{i}))_i=0,
\end{aligned}
$$
where $B_i$ is the $i$-th basis vector of $\mathbb{R}^{3n}$. 
The process starts by setting $q(t_1) = q(t_0) + h / (N - 1)$, and terminates with the result of the exponential map $q(t_{N-1})$. The numerical approach to solve the geodesic initial value problem is implemented into the class \codeobj{DiscreteSurfacesExpSolver}, which is an attribute of the class \texttt{ElasticMetric}.

\subsubsection{Use case: Geodesic between surfaces}

\begin{figure}[h]
    \centering
    \begin{subfigure}[t]{\textwidth}
        \centering
        \includegraphics[width=0.9\textwidth]{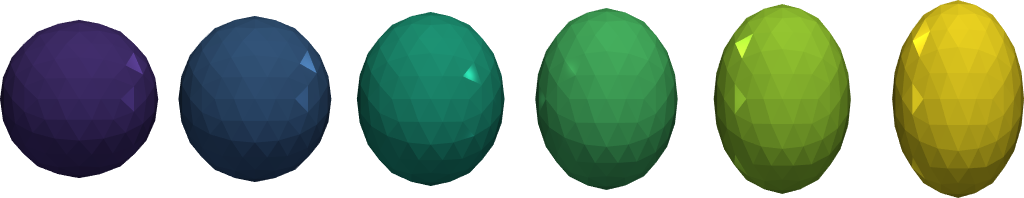}
        \label{fig:geodesic-sphere-ellipsoid}
    \end{subfigure}
    \begin{subfigure}[t]{\textwidth}
        \centering
        \includegraphics[width=0.9\textwidth]{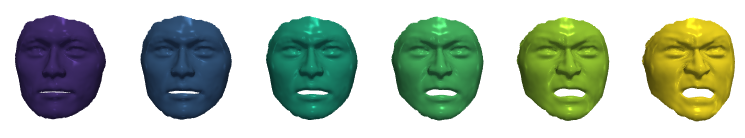}
        \label{fig:geodesic-sphere-faces}
    \end{subfigure}
    \begin{subfigure}[t]{\textwidth}
        \centering
        \includegraphics[width=0.9\textwidth]{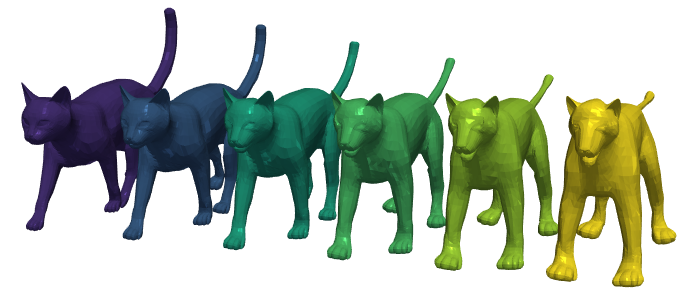}
        \label{fig:geodesic-cat-lion}
    \end{subfigure}
    \caption{\textbf{Deformations of 3D shapes with the \texttt{shape} module of Geomstats.} Geodesics between meshes in the space of discrete surfaces equipped with an elastic metric, with all elastic parameters equal to 1. Top: a sphere deforms into an ellipsoid. Middle: a neutral face deforms into an angry face. Bottom: a cat deforms into a lion. The mesh data of the last two geodesics come from \href{https://github.com/emmanuel-hartman/H2_SurfaceMatch/tree/main/demo}{H2 Surface Match Github Repository}.}
    \label{fig:geodesic-surfaces}
\end{figure}

We show how to perform computations on surface shapes implemented in the \texttt{shape} module using algorithms available in the main Geomstats package. Specifically, we show how to compute geodesics between meshes with known point correspondences (see Figure~\ref{fig:geodesic-surfaces}).

\begin{minted}{python}
from geomstats.geometry.discrete_surfaces import DiscreteSurfaces

# Instantiate space
space = DiscreteSurfaces(faces)

# Compute geodesic
geod_func = space.metric.geodesic(point_a, end_point=point_b)

# Get (chosen) geodesic points
time = gs.linspace(0.0, 1.0, num=6)
geod_points = geod_func(time)

\end{minted}

The geodesics shown in Figure~\ref{fig:geodesic-surfaces} match the geodesics obtained in the original paper \cite{hartman2023elastic}, showing that the Geomstats equivalent geodesic computation is working as intended. Yet, the computation in Geomstats improved the original implementation in that the computation of the inner product defining the metric is substantially faster.

\subsubsection{Opportunities and challenges when computing with surfaces}
While the differential geometric framework for computations with curves and their shapes is relatively mature, the corresponding framework to compute with surfaces and their shapes is newer. For example, the families of elastic Riemannian metric presented in this paper have only been published a year ago~\cite{hartman2023elastic}. Accordingly, both the theory and its numerical implementation contain outstanding challenges and interesting opportunities for research in the field.

First, computations with surfaces rely on numerical algorithms requiring the solution of high-dimensional optimization problems, being inherently slow. This is the case with the computation of the Riemannian logarithm and exponential maps, even when restricted to parameterized surfaces. Dedicated research to improve the speed and accuracy of these approaches would definitely unlock a wider range of applications for the differential geometric branch of shape analysis with surfaces. Second,
the quotient by reparametrizations contains its set of challenges: the question of how to determine whether two triangle meshes represent the same shape is still ill-defined. (Oriented) varifold distances appear to provide an answer in \cite{hartman2023elastic}, but it comes with the additional challenge of tweaking kernels and corresponding parameters. Research investigating the stability of these approaches with respect to the hyperparameters would help advance the field. Besides, the associated high-dimensional optimization problems are even more challenging than their parameterized counterparts. Third, and interestingly, quotienting rotations on the surfaces remains a surprisingly complicated operation when performed jointly with the quotient by reparametrizations. This is due to the fact that there is no definite way of choosing the center of the 3D rotation acting on the surface. One could think of using the barycenter of the surface's vertices. However, reparametrizing the surface changes the position of these points and therefore changes the position of their barycenter. Future theoretical and numerical research should investigate how to perform the joint quotient of rotations and reparameterizations on surfaces. 



\section*{Conclusion}

In this paper, we presented a Python implementation of the differential geometry of shape analysis in the module \texttt{shape} integrated in the software Geomstats. We introduced the architecture of the module and showed that it contains the essential building blocks to perform statistics and machine learning on shape data, through several code snippets. We hope that our implementation will inspire researchers to use, and contribute to, shape analysis with the \codeobj{Geomstats} library.

\section*{Acknowledgements}
This research was partially funded by the National Science Foundation Materials Research Science and Engineering Center (MRSEC) at UC Santa Barbara (NSF DMR-2308708, Data Expert Group and IRG-2). L.F.P. was also partially funded by the MUR under the grant PRIN 2022 project ``GEOPRIDE Geometric primitive fitting on 3D data for geometric analysis and 3D shapes''. This work was also partially funded by the NSF CAREER 2240158. P.S.S. is grateful to the Max-Planck-Institute for Mathematics in Bonn for its hospitality and financial support.


\bibliographystyle{plain}
\bibliography{bibliography}

\end{document}